\documentclass[reprint,aps,prb,amsmath,amssymb,longbibliography, notitlepage]{revtex4-2}
\usepackage{graphicx}
\usepackage{xcolor}
\usepackage{bm}
\usepackage{soul}
\usepackage[normalem]{ulem}
\usepackage[colorlinks, linkcolor= blue, citecolor = blue, urlcolor=blue]{hyperref}
\usepackage{comment}
\usepackage{physics}
\usepackage{textcomp}
\usepackage{pifont}
\usepackage{multirow}
\usepackage{cancel}
\usepackage{wrapfig}
\usepackage[stable]{footmisc}
\usepackage{array}
\usepackage{gensymb}
\usepackage{multirow}
\usepackage{hhline}
\usepackage{booktabs}
\usepackage{braket}
\usepackage{cancel}
\usepackage[colorlinks, linkcolor= blue, citecolor = blue, urlcolor=blue]{hyperref}
\usepackage{chemformula}
\def\nn{\nonumber}
\def\bea{\begin{eqnarray}}
\def\eea{\end{eqnarray}}
\def\be{\begin{equation}}
\def\ee{\end{equation}}

\def\kb{{\bm k}}

\def\e{\varepsilon}

\def\bal{\begin{aligned}}
\def\eal{\end{aligned}}

\newcommand{\cmark}{\ding{51}}%
\newcommand{\xmark}{\ding{55}}%

\begin{document}

\title{Extrinsic orbital Edelstein effect from asymmetric scattering}
\author{Sankar Sarkar}
\email{sankars24@iitk.ac.in}
\author{Koushik Ghorai}
\email{koushikgh20@iitk.ac.in}
\author{Amit Agarwal}
\email{amitag@iitk.ac.in}
\affiliation{Department of Physics, Indian Institute of Technology Kanpur, Kanpur-208016, India}

\begin{abstract}

The generation and manipulation of orbital angular momentum (OAM) by an external electric field constitute one of the central themes of orbitronics. In particular, the electrically induced nonequilibrium OAM polarization, known as the orbital Edelstein effect (OEE), has attracted considerable attention in recent years. While the intrinsic band-geometric mechanism and the role of conventional symmetric impurity scattering in the OEE are well understood, the contribution from disorder-induced asymmetric scattering remains unclear. Here, we develop a semiclassical theory that separates the OEE into intrinsic, Drude, side-jump, and third- and fourth-order skew-scattering channels. Unlike the Drude channel, the intrinsic, side-jump, and skew-scattering responses survive only in systems with broken time-reversal symmetry. We find that in a magnetized Rashba two-dimensional electron gas (2DEG), these disorder-induced mechanisms can substantially exceed the intrinsic contribution. Remarkably, we find that for a system with Rashba coupling of $1~\mathrm{eV\,\text{\AA}}$, the orbital magnetization is about one order of magnitude larger than the spin magnetization for the chosen parameters, highlighting the crucial role of orbital degrees of freedom in the Edelstein effect.

\end{abstract}

\maketitle

\section{Introduction}

Recently, orbitronics~\cite{Go_2017_scirep, Go_2021_epl, Choi_2023_nature, Lyalin_2023_prl} has emerged as a promising platform for electrical generation and manipulation of magnetization~\cite{Niu_07_prl, Go_2020_prr, Lee_2021_natcomm, Manchon_2024_ncomm} by harnessing the orbital degree of freedom of electrons \cite{Bernevig_2005_prl, Go_2018_prl, Bhowal_2020_prb, Kamal_21_prb, Rhonald_2024, Koushik_2026_gyro, Koushik_2026_splitter, Sunit_26_arxiv}. Two primary channels for generating a nonequilibrium OAM density are the OEE~\cite{Yoda_2018_nanolett, Johansson_2021, Johansson_2023_prr, Mertig_25_prb, Murakami_2020_prb, Witt_2026} and the orbital Hall effect (OHE)~\cite{Bernevig_2005_prl, Go_2018_prl, Bhowal_2020_prb, Bansil_2026_rop,Sarkar_2026_orbital}. The OEE generates a net magnetization in noncentrosymmetric systems through electric-field-induced redistribution of occupations of OAM-textured Bloch states. The OHE, on the other hand, generates a transverse flow of orbital angular momentum, leading to OAM accumulation at the sample edges, and can occur in both centrosymmetric and noncentrosymmetric systems. Microscopically, the OHE originates from intrinsic band geometry~\cite{Salemi_2022_prm, Dimi_2025_prl} and extrinsic skew-scattering and side-jump mechanisms~\cite{Tang_2024_prl, Rappoport_2025_prl}. In particular, analogous to the charge~\cite{ZZDu_19_Natcomm, Atencia_2023_prb, Pesin_17_prb, Harsh_2026_prb} and spin Hall effects~\cite{Mertig_2010_prl, CastroNeto_2014_prb, Sinova_2015_RMP, Fujimoto_2017_jps, Ando_2019_prm, sanjay_2026_prb}, recent work has shown that extrinsic contributions can substantially dominate the intrinsic OHE in two-dimensional massive Dirac systems~\cite{Liu_2024_prl, Cong_Xiao_26}. Despite these advances, existing theories of the OEE remain limited to symmetric impurity scattering within the relaxation-time approximation and intrinsic band-geometric mechanisms~\cite{Murakami_15_scirep, Johansson_2021, Johansson_2023_prr, Johansson_2024_jpc, Witt_2026, Manchon_26_prb, Dimi_25_arxiv, XCXie_26_apl}, leaving the role of asymmetric scattering largely unexplored. This raises a natural question: how do side-jump and skew-scattering mechanisms modify the OEE?

\begin{figure}[t!]
    \centering
    \includegraphics[width=\linewidth]{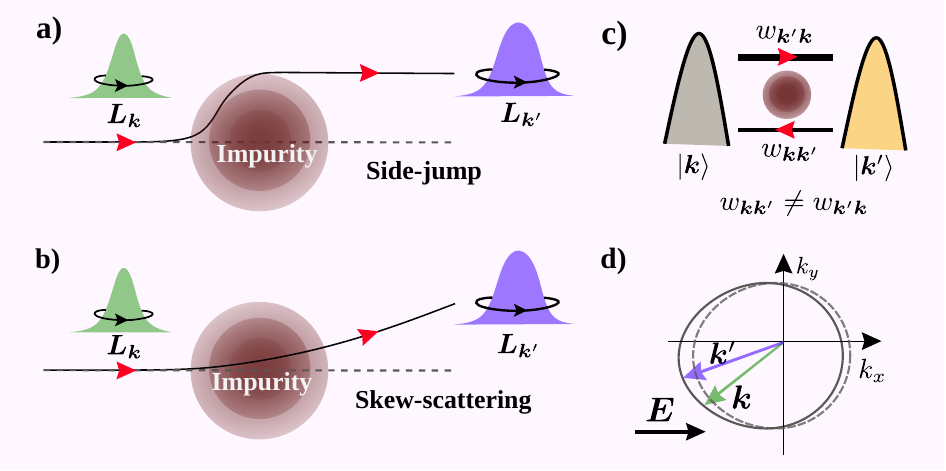}
    \caption{\textbf{Schematic of the disorder-induced contributions to the OEE.} (a) An electron with initial OAM $\bm{L}_{\bm{k}}$ undergoes impurity scattering, acquires a side-jump coordinate shift, and is scattered into a state with OAM $\bm{L}_{\bm{k}'}$. (b) Asymmetric skew scattering bends the electron trajectory, connecting states with OAM $\bm{L}_{\bm{k}}$ and $\bm{L}_{\bm{k}'}$. (c) Asymmetric scattering is characterized by unequal forward and backward scattering rates, $w_{\bm{k}\bm{k}'} \neq w_{\bm{k}'\bm{k}}$, between states with different momenta. (d) The combination of asymmetric scattering and the electric-field-shifted Fermi surface generates the disorder-driven orbital magnetization. The dotted circle in (d) represents the unperturbed Fermi surface. For simplicity, we drop the band indices.}
    \label{Fig:Fig_Schematic}
\end{figure}

To answer this question, we develop a microscopic semiclassical Boltzmann theory of the OEE beyond the conventional relaxation-time approximation. In addition to the intrinsic band-geometric and conventional symmetric-scattering contributions, we identify two disorder-induced mechanisms: (i) a side-jump contribution arising from the field-induced coordinate shift of Bloch electrons and (ii) skew-scattering contributions originating from antisymmetric impurity scattering (see Fig.~\ref{Fig:Fig_Schematic}). Apart from the conventional OEE, these extrinsic contributions are odd under time reversal ($\mathcal{T}$) and therefore arise only in noncentrosymmetric magnetic systems. Guided by the magnetic point-group symmetry analysis, we identify a magnetized Rashba 2DEG with both in-plane and out-of-plane exchange-field components as a minimal model to demonstrate this response. The in-plane component breaks the continuous rotational symmetry, while the out-of-plane component opens a gap at the shifted Rashba band crossing. Near this avoided band-crossing, the Bloch-state OAM reaches approximately \(7\hbar\) for the chosen parameters. The resulting OEE response substantially exceeds its spin counterpart and is dominated by the side-jump and skew-scattering contributions. Finally, we show that the orbital magnetization can be efficiently controlled by tuning both the exchange field and the Rashba spin--orbit coupling (SOC). In particular, the OEE exhibits a pronounced dependence on the magnetization orientation and scales quadratically with the Rashba SOC strength, whereas the spin Edelstein effect nearly saturates. For the selected parameters, our numerical results show that the orbital response can be an order of magnitude  larger than its spin counterpart.

The remainder of the paper is organized as follows. In Sec.~\ref{sec:Theory}, we develop a semiclassical Boltzmann framework for the OEE. The effects of the electric field and disorder on the OAM and distribution function, and their subsequent contributions to the total orbital polarization, are discussed in separate subsections. In Sec.~\ref{sec:symmetry_analysis}, we analyze the fundamental and crystallographic symmetry constraints on the different orbital susceptibility tensors. In Sec.~\ref{sec:Rashba_2DEG}, we apply the theory to a magnetized Rashba 2DEG and examine the chemical-potential dependence of the OEE. We further explore the tuning of the orbital response through the exchange-field direction and Rashba spin--orbit coupling and compare the orbital and spin contributions to the current-induced magnetization in separate subsections. We conclude and summarize our findings in Sec.~\ref{sec:conclusion}.


\section{Theory of the extrinsic orbital Edelstein effect} \label{sec:Theory}


In this section, we develop a semiclassical theory for extrinsic OEE (EOEE) that includes several disorder-induced channels. Here, we consider the effects of the driving field and disorder on equal footing. Assuming that both the external electric field and the disorder potential are sufficiently weak, they can be treated as perturbations to the bare crystal Hamiltonian, $\mathcal{H}$. The effects of these perturbations are twofold: first, they modify the Bloch states and accordingly induce corrections to the OAM; second, the perturbations redistribute the carriers in momentum space. Within the semiclassical framework, the total OAM polarization can be expressed as \cite{Johansson_2021, Johansson_2023_prr, Cong_Xiao_24_prb, sanjay_2026_prb}
\begin{align}
    \delta L^{\nu}
    = - \frac{\mu_B }{\hbar} \sum_l \tilde{L}_l^{\nu} f_l ~.\label{eq: def_OEE}
\end{align}   
Here, $\mu_B$ is the Bohr magneton, $\hbar$ is the reduced Planck constant, $\tilde{L}^{\nu}_l = \braket{\tilde{l} | \hat{L}^{\nu} | \tilde{l}}$ is the OAM carried by the perturbed state $\ket{\tilde{l}}=\ket{l}+\ket{\delta l_{\bm E}}+\ket{\delta l_{\rm dis}}$ along $\nu~(=x,y,z)$ direction and $f_l$ is the nonequilibrium Fermi--Dirac distribution function. The Bloch state is $\ket{l} = e^{i \bm k \cdot \bm r} \ket{u_{n\bm k}}$, where $\ket{u_{n\bm k}}$ denotes its cell-periodic part. The composite index $l \equiv (n,\bm{k})$ labels the band index $n$ and crystal momentum $\bm{k}$. In the perturbed state, $\ket{\delta l_{\bm E}}$ and $\ket{\delta l_{\rm dis}}$ are the corrections due to the electric field ($\bm E$) and disorder potential, respectively. We derive the perturbed OAM in Sec.~\ref{Modified_OAM}, the field- and disorder-corrected distribution function in Sec.~\ref{Non_eq_distribution}, and the resulting orbital magnetization in Sec.~\ref{response_tensors}. In Eq.~\eqref{eq: def_OEE}, we use the shorthand notation $\sum_l \equiv \sum_n \int d^Dk/(2\pi)^D,$ where $D$ is the spatial dimension of the system.


\subsection{Field- and disorder-corrected orbital angular momentum}\label{Modified_OAM}


The field- and disorder-dressed OAM is given by $\tilde{L}^{\nu}_l = \braket{\tilde{l} | \hat{L}^{\nu} | \tilde{l}}$. Based on their physical origins, the total OAM can be decomposed into three contributions,
\begin{align}
    \tilde{L}^{\nu}_l = L_l^{\nu} + L_l^{\nu,\mathrm{anm}} + L_l^{\nu,\mathrm{sj}}~.
    \label{eq: orbital_angular_momentum}
\end{align}
Here, $L^{\nu}_l=\braket{l|\hat{L}^{\nu}|l}$ is the OAM of the unperturbed Bloch state, while $L^{\nu,\mathrm{anm}}_l$ and $L^{\nu,\mathrm{sj}}_l$ arise from the electric-field- and disorder-induced corrections to the Bloch state, corresponding to the anomalous and side-jump contributions, respectively. The matrix element of the unperturbed OAM is given by~\cite{Cysne_2026_prr, Dongwook_2026_prb}
\begin{align}
    \bm L_{nm} =- \frac{ie\hbar^2}{4 g_L \mu_B} \Bigg( \sum_{p \neq n} \frac{\bm v_{np} \times \bm v_{pm}}{\e_{p\kb} - \e_{n\kb}} + \sum_{p \neq m} \frac{\bm v_{np} \times \bm v_{pm}}{\e_{p\kb} - \e_{m\kb}} \Bigg)~.\label{OAM_matrix_elements}
\end{align}
Here, $-e$ ($e>0$) is the electronic charge and  $g_L \approx 1$ is the orbital $g$ factor. The velocity matrix elements are defined as $\bm{v}_{nm}=\bra{u_{n\bm{k}}}\hat{\bm{v}}\ket{u_{m\bm{k}}}$, where $\hat{\bm{v}}=(1/\hbar)\partial_{\bm{k}}\mathcal{H}$. Throughout this work, we use the notation $X_{nm}=\bra{u_{n\bm{k}}}\hat{X}\ket{u_{m\bm{k}}}$ to represent the matrix elements of any operator $\hat{X}$ in the Bloch basis. The anomalous contribution to OAM is given by
\begin{align}
    L_{l}^{\nu,\mathrm{anm}}
    = 2 \Re \braket{l| \hat{L}^{\nu}|\delta l_{\bm E}} =
    -\frac{e}{\hbar}\,
    \mathcal{U}_{l}^{\nu,a}E_a~.\label{eq: anomalous_OAM}
\end{align}
In analogy with the anomalous spin polarizability \cite{SARKAR_2026_MTQ, Cong_xiao_2023_prl, Cong_Xiao_24_prb, sanjay_2026_prb}, we define the band-geometric quantity
$\mathcal{U}_{n\bm{k}}^{\nu,a}$ as the \emph{anomalous orbital polarizability} (AOP)~\cite{Shengyuan_24_prl},
\begin{align}
    \mathcal{U}_{n\bm{k}}^{\nu,a} = -2\hbar^2 \Im \sum_{m\neq n} \frac{ L^{\nu}_{nm} v^a_{mn}}{(\varepsilon_{n\bm{k}}-\varepsilon_{m\bm{k}})^2}~.
\end{align}

To evaluate the disorder-induced side-jump contribution to the OAM, we parameterize the disorder cumulants using the randomly distributed delta-function potential $V_{\mathrm{imp}}(\bm r)=\sum_i V_i\,\delta(\bm r-\bm R_i),$ where $V_i$ and $\bm R_i$ denote the impurity potential strength and position, respectively. The side-jump contribution to the OAM originates
from the second-order symmetric scattering rate and is given by 
\begin{widetext} 
\begin{align}
    L_l^{\nu,\text{sj}} &= - 2 \pi \sum_{n', \bm k'} W_{\bm k \bm k'} \delta( \varepsilon_{n \bm k} - \varepsilon_{n' \bm k'} )
    \times \text{Im} \Bigg[ \sum_{n'' \neq n'} \frac{L^{\nu}_{n' n''} (\bm k')   u^{\bm k \bm k'}_{n n'} u^{\bm k' \bm k}_{n'' n} }{(\varepsilon_{n' \bm k'} - \varepsilon_{n'' \bm k'})} -\sum_{n '' \neq n} \frac{L^{\nu}_{n n''} (\bm k)  u^{\bm k \bm k'}_{n'' n'} u^{\bm k' \bm k}_{n' n} }{(\varepsilon_{n \bm k} - \varepsilon_{n'' \bm k}) } \Bigg]~.\label{eq: side_jump_e1}
\end{align}
\end{widetext}
The detailed calculation of $L_l^{\nu,\text{sj}}$ is presented in Appendix~\ref{side_jump_L}. For convenience, we define the overlap between the cell-periodic parts of the Bloch states as $u^{\bm k \bm k'}_{nn'}=\braket{u_{n\bm k}|u_{n'\bm k'}}$. Within the first Born approximation, the second-order scattering amplitude for static, randomly distributed impurities is defined as $W_{\bm k \bm k'} = \langle \left|V^{0}_{\bm k \bm k'}\right|^2 \rangle_{\mathrm{dis}}$, where $\langle\cdots\rangle_{\mathrm{dis}}$ denotes the disorder average. Carrying out this average restores translational symmetry at the disorder-averaged level, allowing us to work in the crystal-momentum representation. Within an intraband small-momentum-transfer approximation, where $\bm q=\bm k'-\bm k\rightarrow0$ and $n'=n$ for the DC response, the side-jump contribution simplifies to \cite{Cong_Xiao_24_prb}
\begin{align}
    L_l^{\nu, \rm sj} = \frac{2 \pi}{\hbar} \sum_{\bm k'} W_{\bm k\bm k'} \delta(\varepsilon_{n\bm k} - \varepsilon_{n\bm k'}) [(k_a - k_a')\,\mathcal{U}^{\nu,a}_l]~.
\end{align}
Notably, both the anomalous and side-jump corrections originate from interband coherence and are governed by the AOP. Together, these corrections and the conventional OAM give the total OAM of a Bloch electron. The electric field and impurity scattering also drive the electron distribution out of equilibrium. We derive the resulting nonequilibrium distribution function in the following section.


\subsection{Non-equilibrium distribution function}\label{Non_eq_distribution}


To account for the effects of the impurity potential $V_{\mathrm{imp}}$ and the electric field on the distribution function, we employ the semiclassical Boltzmann equation. Assuming elastic impurity scattering and a spatially uniform electric field, the nonequilibrium distribution function evolves according to \cite{Harsh_2026_prb, sanjay_2026_prb}
\begin{align}
    \frac{\partial f_l}{\partial t} + \dot{\bm k}\cdot\partial_{\bm k}f_l = I_{\mathrm{el}}\{f_l\}~.
\end{align}
The collision integral $I_{\mathrm{el}}\{f_l\}$ reflects the elastic scattering processes between Bloch states \cite{Fu_2021_prb},
\begin{align}
    I_{\mathrm{el}}\{f_l\} = -\sum_{l'} \left( w_{l'l}f_l - w_{ll'}f_{l'} \right)~.\label{eq:collision_integral}
\end{align}
Here, $w_{ll'}$ denotes the scattering rate from state $\ket{l'}$ to state $\ket{l}$. This rate is determined by Fermi's golden rule \cite{sakurai2020modern},
\begin{align}
    w_{ll'} = \frac{2\pi}{\hbar} \left\langle
    \left| \braket{l|V_{\mathrm{imp}}|l'_{\mathrm{dis}}} \right|^2 \right\rangle_{\mathrm{dis}} \delta(\varepsilon_l-\varepsilon_{l'})~.
\end{align}
The delta function $\delta(\varepsilon_l - \varepsilon_{l'})$ ensures energy conservation between the scattering states. Here, $\ket{l'_{\text{dis}}}$ is the eigenstate of the full Hamiltonian, $ H = \mathcal{H} + V_{\text{imp}} $, which satisfies the Lippmann--Schwinger equation \cite{sakurai2020modern},
\begin{align}
    \ket{l_{\text{dis}}} = \ket{l} + (\varepsilon_l - \mathcal{H} + i\eta)^{-1} V_{\text{imp}} \ket{l_{\text{dis}}}~.
\end{align}
The infinitesimal parameter $i\eta \rightarrow 0^+$ enforces the outgoing boundary condition for the scattering states.

In general, higher-order scattering rates are not symmetric under the exchange of the initial and final states, i.e., $ w_{ll'} \neq w_{l'l}.$ It is therefore convenient to decompose the scattering rate into its symmetric and antisymmetric components and analyze their respective contributions to transport. Accordingly, we write the scattering rate as $w_{ll'} = w_{ll'}^{\rm S} + w_{ll'}^{\rm A},$
where
\begin{equation}
w_{ll'}^\mathrm{S} = w_{l'l}^{\mathrm{S}} = \frac{w_{ll'}+w_{l'l}}{2}~,
\quad
w_{ll'}^\mathrm{A} = -w_{l'l}^\mathrm{A} = \frac{w_{ll'}-w_{l'l}}{2}~.\nn
\end{equation}
The symmetric component, $w_{ll'}^{\mathrm S}$, describes the conventional relaxation processes and is commonly treated within the relaxation-time approximation. Most of the previous studies of OEE have been restricted to this relaxation-time approximation, and have obtained only the Drude-like Fermi-surface response \cite{Johansson_2021, Johansson_2023_prr, Witt_2026, Gautam_2026_prb}. However, the relaxation-time approximation does not capture the effect of asymmetric disorder scattering. The antisymmetric component, $w_{ll'}^{\rm A}$, describes such asymmetric scattering and gives rise to skew-scattering contributions to the OEE, which we discuss in the next section.

To systematically account for the various scattering mechanisms, we consider the weak-disorder limit and expand the scattering rates in powers of the impurity potential strength,
\begin{align} \label{w_ll_expansion}
    w_{ll'} = w_{ll'}^{(2)} + w_{ll'}^{(3),\mathrm{A}} + w_{ll'}^{(4),\mathrm{A}}~.
\end{align}
Here, $w_{ll'}^{(3),\mathrm{A}}$ and $w_{ll'}^{(4),\mathrm{A}}$ denote the antisymmetric scattering rates arising at third and fourth order in the impurity potential, respectively. The corresponding symmetric contributions are not written explicitly, as they merely renormalise the second-order symmetric scattering rate and can therefore be absorbed into $w_{ll'}^{(2),\mathrm{S}}$. The second-order contribution can be further separated into a field-independent symmetric term and an electric-field-induced coordinate-shift contribution \cite{ZZDu_19_Natcomm, Harsh_2026_prb},
\begin{align}
    w_{l'l}^{(2),\mathrm{S}} &= \frac{2\pi}{\hbar} \langle |V_{l'l}|^2\rangle_{\rm dis}
    \delta(\varepsilon_l-\varepsilon_{l'})~, \\
    w_{l'l}^{(2),\rm cs} &=- \frac{2\pi}{\hbar} \langle |V_{l'l}|^2\rangle_{\rm dis}
    \frac{\partial\delta(\varepsilon_l-\varepsilon_{l'})}{\partial\varepsilon_l} e\bm E\cdot\delta\bm r_{l'l}~.
\end{align}
The coordinate shift $\delta \bm r_{l'l}$ of the electronic wave packet during a scattering event is given by~\cite{MacDonald_2006_prb}
\begin{equation}
\label{appx:coordinate_shift}
\delta \bm r_{l'l} = \braket{u_{n'\bm k'}|i\partial_{\bm k'}|u_{n'\bm k'}}
- \braket{u_{n\bm k}|i\partial_{\bm k}|u_{n\bm k}} -
D_{\kb\kb'} \arg(V_{l'l})~.
\end{equation}
Here, we define $D_{\bm{k}\bm{k}'} = \left(\partial_{\bm{k}} + \partial_{\bm{k}'}\right)$, and $\arg$ denotes the phase of a complex number. The detailed derivations of the scattering rates are presented in Appendix \ref{Simplification_of_scattering_rates}.

Substituting the scattering-rate expansion in Eq.~(\ref{w_ll_expansion}) into Eq.~(\ref{eq:collision_integral}), we separate the collision integral according to the underlying scattering mechanisms,
\begin{equation}
I_{\rm el}\{f_l\} = I_{\rm el}^{\rm Dr}\{f_l\} + I_{\rm el}^{\rm sj}\{f_l\} +
I_{\rm el}^{\rm sk3}\{f_l\} + I_{\rm el}^{\rm sk4}\{f_l\}~.\nn
\end{equation}
The Drude (``Dr"), $I_{\rm el}^{\rm Dr}\{f_l\}$, and side-jump (``sj"), $I_{\rm el}^{\rm sj}\{f_l\}$, contributions arise from the symmetric and coordinate-shift parts of the second-order scattering rate, respectively. The skew-scattering (``sk") contributions, $I_{\rm el}^{\rm sk3}\{f_l\}$ and $I_{\rm el}^{\rm sk4}\{f_l\}$, originate from the antisymmetric parts of the third- and fourth-order scattering rates, respectively. The explicit expressions for these contributions are
\begin{align}
    &I_{\rm el}^{\rm Dr}\{f_l\}  = -\sum_{l'} w_{ll'}^{(2),\mathrm{S}} (f_l-f_{l'})~,\\
    &I_{\rm el}^{\rm sj}\{f_l\}  = -\sum_{l'} w_{ll'}^{(2),\rm cs} (f_l-f_{l'})~, \\
    &I_{\rm el}^{\rm sk3}\{f_l\} = \sum_{l'} w_{ll'}^{(3),\rm A} (f_l+f_{l'})~, \\
    &I_{\rm el}^{\rm sk4}\{f_l\} = \sum_{l'} w_{ll'}^{(4),\rm A} (f_l+f_{l'})~.
\end{align}
The nonequilibrium distribution function can likewise be decomposed according to the underlying scattering mechanisms,
\begin{equation}
f_l = f_l^{\rm Dr} + f_l^{\rm sj} + f_l^{\rm sk3} + f_l^{\rm sk4}~.
\end{equation}
We substitute this decomposition into the Boltzmann equation. For a time-independent electric field in the steady state, $\partial f_l/\partial t = 0$, and the electron dynamics are governed by $\dot{\bm k} = -e\bm E/\hbar$. Solving the resulting coupled Boltzmann equations to linear order in the electric field, we obtain the first-order corrections to the distribution function as
\begin{align}
&f_l^{\rm Dr,(1)} = \frac{e\tau}{\hbar} \,\bm E\cdot\partial_{\bm k}f_l^0~,\\
&f_l^{\rm sj,(1)} = -e\tau\, \bm E\cdot\bm v_l^{\rm sj} \frac{\partial f^0} {\partial\varepsilon_l}~, \\
&f_l^{\rm sk3,(1)} = \frac{e\tau^2}{\hbar} \sum_{l'} w_{ll'}^{(3),\rm A} \left(
\bm E\cdot\partial_{\bm k}f_l^0 + \bm E\cdot\partial_{\bm k}f_{l'}^0 \right)~, \\
&f_l^{\rm sk4,(1)} = \frac{e\tau^2}{\hbar} \sum_{l'} w_{ll'}^{(4),\rm A} \left( \bm E\cdot\partial_{\bm k}f_l^0 + \bm E\cdot\partial_{\bm k}f_{l'}^0 \right)~.
\end{align}
The detailed derivation of the distribution function is presented in Appendix~\ref{non_equilibrium_distributiom}. Here, $\tau$ denotes the relaxation time, determined by the symmetric second-order scattering rate $w_{ll'}^{(2),\mathrm{S}}$. The equilibrium distribution is given by $f^0_l = [1 + \exp\{(\varepsilon_l - \mu)/k_B T\}]^{-1}$, where $k_B$ is the Boltzmann constant and $T$ and $\mu$ denote the equilibrium temperature and chemical potential, respectively. In the side-jump contribution $f_l^{\mathrm{sj},(1)}$, the side-jump velocity is defined as~\cite{MacDonald_2006_prb,sanjay_2026_prb}
\begin{equation}
\label{sj_vel}
\bm v_l^{\rm sj} = \sum_{l'} w_{l'l}^{(2),\rm S} \,\delta\bm r_{l'l}~.
\end{equation}
The detailed expressions for the third- and fourth-order antisymmetric scattering rates and the side-jump velocity are provided in Appendices \ref{app_third_order_scattering_rates}, \ref{app_fourth_order_scattering_rates}, and \ref{side_jump_velocity}, respectively. For completeness, we present here the corresponding simplified expressions:

\begin{align}
    w^{(3),\rm A}_{n,\bm k\bm k'} &= -\frac{2\pi^2 n_i V_1^3}{\hbar}\sum_{ \kb''} \delta(\varepsilon_{ n \bm k}- \varepsilon_{ n \bm k'})\delta(\varepsilon_{ n\bm k} - \varepsilon_{ n \bm k''}) \nn \\ 
    &\quad  \times [(\kb''\times\kb')+(\kb'\times\bm{k})+(\bm{k}\times\kb'')]\cdot\bm{\Omega}_ n(\bm{k}) ~, \\
    w^{(4),\rm A}_{n,\bm k\bm k'} &= -\frac{2\pi^2 n_i^2 V_0^4}{\hbar}\sum_{ \kb''} \delta(\varepsilon_{ n \bm k}- \varepsilon_{ n \bm k'})\delta(\varepsilon_{ n\bm k} - \varepsilon_{ n \bm k''}) \nn \\ 
    &\quad \times [(\kb''\times\kb')+(\kb'\times\bm{k})+(\bm{k}\times\kb'')]\cdot\tilde{\bm{\Omega}}_ n(\bm{k}) ~,\\
    \bm v^{\rm sj}_{n\bm k} &= \frac{2 \pi n_i V_0^2}{\hbar} \sum_{\bm k'} [ (\bm k - \bm k') \times \mathbf{\Omega}_n(\bm k) ] \delta(\varepsilon_{n\bm k} - \varepsilon_{n\bm k'})~.
\end{align}
Here, $n_i$, $V_0$, and $V_1$ denote the impurity density, zeroth-order moment, and first-order moment of the impurity potential, respectively. Moreover, $\mathbf{\Omega}_{n}(\bm{k})$ and $\tilde{\mathbf{\Omega}}_{n}(\bm{k}) = \sum_{n' \neq n} \mathbf{\Omega}_{nn'}(\bm{k})/(\varepsilon_{n\bm{k}} - \varepsilon_{n'\bm{k}})$ denote the Berry curvature and energy-normalized Berry curvature of band $n$, respectively.


\subsection{Components of the orbital susceptibility tensor}\label{response_tensors}


\begin{figure*}[t!]
\centering
\includegraphics[width=0.85\linewidth]{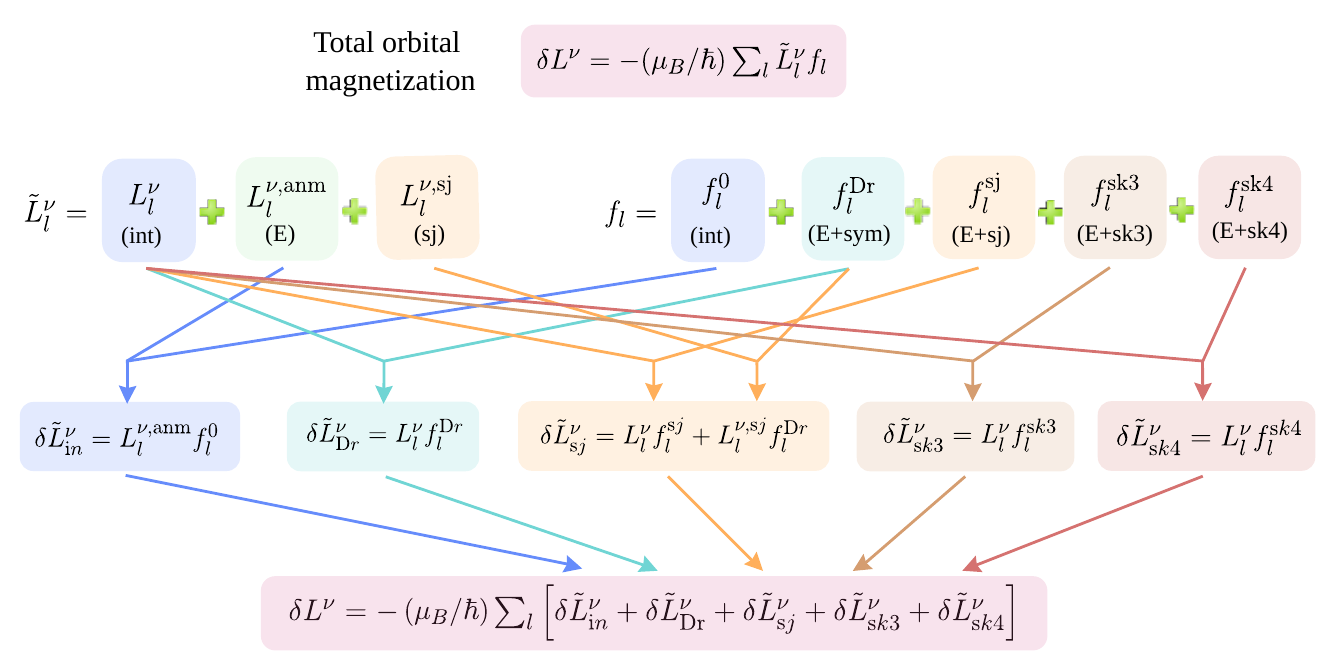}
\caption{\textbf{Components of OEE.}
The total orbital magnetization is determined by the electric field ($E$)- and disorder-dressed OAM and the nonequilibrium distribution function. The dressed OAM consists of the unperturbed contribution (int), the field-induced anomalous correction (anm), and the side-jump correction (sj). The distribution function is decomposed into the equilibrium contribution (int) and four nonequilibrium ($\sim E$) components according to the underlying scattering mechanisms: the conventional Drude contribution (Dr) from symmetric scattering (sym), the side-jump contribution (sj), and the third- and fourth-order skew-scattering contributions (sk3 and sk4). Combining these contributions yields five distinct channels of the OEE: intrinsic, Drude, side-jump, third-order skew-scattering, and fourth-order skew-scattering. The side-jump and skew-scattering channels constitute the disorder-induced novel contributions to the OEE.}
\label{Fig:Fig_OEE_origin_tree}
\end{figure*} 

Substituting the field- and disorder-modified OAM and first-order distribution function into Eq.~(\ref{eq: def_OEE}), we obtain the linear-order orbital polarization,
\begin{align}
    \delta L^{\nu} &= -\frac{\mu_B}{\hbar}  \sum_l \big[ L^{\nu}_l ( f_l^{\rm Dr, (1)} + f_l^{\rm sj, (1)} + f_l^{\rm sk3, (1)} + f_l^{\rm sk4, (1)} ) \nn \\
    &\qquad\qquad+ L_l^{\nu, \rm anm} f_l^0 + L_l^{\nu, \rm sj} f_l^{\rm Dr ,(1)} \big] \nn \\
    &= \delta L^{\nu}_{\rm in} + \delta L^{\nu}_{\rm Dr} + \delta L^{\nu}_{\rm sj} + \delta L^{\nu}_{\rm sk3} + \delta L^{\nu}_{\rm sk4}~.
\end{align}
%
According to their physical origins, the total orbital polarization naturally decomposes into five contributions: (i) the intrinsic contribution, $\delta L^{\nu}_{\rm in}$, arising from the field-induced anomalous OAM ($L_l^{\nu,\mathrm{anm}}$) and governed by the AOP; (ii) the Drude or conventional contribution, $\delta L^{\nu}_{\rm Dr}$, arising from the usual relaxation-time approximation; (iii) the side-jump contribution, $\delta L^{\nu}_{\rm sj}$, arising from the side-jump correction to the OAM and distribution function; and (iv) and (v) the skew-scattering contributions, $\delta L^{\nu}_{\rm sk3}$ and $\delta L^{\nu}_{\rm sk4}$, arising from the third- and fourth-order antisymmetric scattering rates, respectively. The explicit expressions for these contributions are
%
\begin{align}
    &\delta L^{\nu}_{\rm in} = -\frac{\mu_B}{\hbar} \sum_l L^{\nu, \rm anm}_l f_l^0 ~,\\
    &\delta L^{\nu}_{\rm Dr} = -\frac{\mu_B}{\hbar} \sum_l L^{\nu}_l f_l^{\rm Dr, (1)}~, \\
    &\delta L^{\nu}_{\rm sj} = -\frac{\mu_B}{\hbar} \sum_l (L^{\nu}_l f_l^{\rm sj, (1)} + L_l^{\nu, \rm sj} f_l^{\rm Dr, (1)} )~,\\
    &\delta L^{\nu}_{\rm sk3} = -\frac{\mu_B}{\hbar} \sum_l L^{\nu}_l f_l^{\rm sk3, (1)}~,\\
    &\delta L^{\nu}_{\rm sk4} = -\frac{\mu_B}{\hbar} \sum_l L^{\nu}_l f_l^{\rm sk4, (1)}~.
\end{align}
For the linear OEE, these contributions can be expressed in terms of the corresponding orbital susceptibility tensors as $ \delta L^{\nu} =  (\chi^{\nu, a}_{\rm in} + \chi^{\nu, a}_{\rm Dr} + \chi^{\nu, a}_{\rm sj} + \chi^{\nu, a}_{\rm sk3} + \chi^{\nu, a}_{\rm sk4}) E_a $, where the individual susceptibility components are
\begin{align}
    \chi^{\nu, a}_{\rm in} &= \frac{e\mu_B}{\hbar^2} \sum_l \mathcal{U}^{\nu,a}_l f_l^0 ~, \label{chi_intrinsic}\\
    \chi^{\nu, a}_{\rm Dr} &= -\frac{e\mu_B\tau}{\hbar} \sum_l L^{\nu}_l v^a_l \frac{\partial f_l^0}{\partial \varepsilon_l}~, \label{chi_Dr}\\
    \chi^{\nu, a}_{\rm sj} &=- \frac{e \mu_B \tau}{\hbar} \sum_l \Big( L^{\nu, \rm sj}_l v_l^a -  L^{\nu}_l v^{\text{sj},a}_l  \Big) \frac{\partial f_l^0}{\partial \varepsilon_l}~,\label{chi_sj}\\
    \chi^{\nu,a}_{\rm sk3} &=- \frac{e \mu_B \tau^2}{\hbar} \sum_{l,l'} w^{(3),\rm A}_{ll'} (L^{\nu}_l - L^{\nu}_{l'}) v_l^a \frac{\partial f_l^0}{\partial \varepsilon_l}~,\label{chi_sk3} \\
    \chi^{\nu,a}_{\rm sk4} &=- \frac{e \mu_B \tau^2}{\hbar} \sum_{l,l'} w^{(4),\rm A}_{ll'} (L^{\nu}_l - L^{\nu}_{l'}) v_l^a \frac{\partial f_l^0}{\partial \varepsilon_l}~\label{chi_sk4}.
\end{align}
These five contributions are the dominant response channels within the weak-disorder approximation. Their origin is summarized schematically in Fig.~\ref{Fig:Fig_OEE_origin_tree}. The explicit factors of $\tau$ in Eqs.~(\ref{chi_Dr})--(\ref{chi_sk4}) do not by themselves give the net disorder scaling. For $w^{(2),\mathrm S}\propto n_iV_0^2$, one has $\tau\propto(n_iV_0^2)^{-1}$ and $L^{\mathrm{sj}},v^{\mathrm{sj}}\propto n_iV_0^2$. Thus, at fixed disorder amplitudes, $\chi_{\mathrm{Dr}}\propto(n_iV_0^2)^{-1}$, whereas $\chi_{\mathrm{sj}}\propto\tau^0$. Similarly, $\chi_{\mathrm{sk3}}\propto\tau^2n_iV_1^3$ and $\chi_{\mathrm{sk4}}\propto\tau^2n_i^2V_0^4\propto\tau^0$. Notably, the Drude, side-jump, and skew-scattering terms are all Fermi-surface responses, as they involve $\partial f_l^0/\partial\varepsilon_l$. In contrast, the intrinsic contribution is a Fermi-sea response and can therefore appear in both metallic and insulating systems.


\section{Symmetry restrictions on response tensors} \label{sec:symmetry_analysis}

%
\renewcommand{\arraystretch}{1.5}
\begin{table}[t!]
\caption{Transformation properties of momentum-dependent physical quantities under inversion ($\mathcal{P}$) and time reversal ($\mathcal{T}$). Here, $\cancel{\mathcal{P}}$ and $\cancel{\mathcal{T}}$ denote broken space-inversion and time-reversal symmetries, respectively.}
\centering
\begin{tabular*}{\columnwidth}{@{\extracolsep{\fill}} c c c}
\hline\hline
Quantities &
$\mathcal{P},\,\cancel{\mathcal{T}}$ &
$\cancel{\mathcal{P}},\,\mathcal{T}$ \\
\hline
$\kb$ & $-\kb$ &  $-\kb$\\
$\e_n (\kb)$ & $\e_n (-\kb)$ & $\e_n (-\kb)$ \\
$\bm v_n (\kb)$ & $ - \bm v_n (-\kb)$ & $-\bm v_n (-\kb)$ \\
$L^{\nu, 0}_n (\kb)$ & $L^{\nu, 0}_n (-\kb)$ & $-L^{\nu, 0}_n (-\kb)$ \\
$\bm \Omega_n (\kb)$ & $\bm \Omega_n (-\kb)$ & $-\bm \Omega_n (-\kb)$ \\
$\mathcal{U}^{\nu,a}_n (\kb)$ & $-\mathcal{U}^{\nu,a}_n (-\kb)$ & $-\mathcal{U}^{\nu,a}_n (-\kb)$ \\
$\bm v^{\rm sj}_n (\kb)$ & $-\bm v^{\rm sj}_n (-\kb)$ & $\bm v^{\rm sj}_n (-\kb)$ \\
$L^{\nu; \rm sj}_n (\kb)$ & $L^{\nu; \rm sj}_n (-\kb)$ & $L^{\nu; \rm sj}_n (-\kb)$ \\
$w^{(3/4), \rm A}_n (\kb, \kb')$ & $w^{(3/4), \rm A}_n (-\kb, -\kb')$ & $-w^{(3/4), \rm A}_n (-\kb, -\kb')$ \\
\noalign{\vskip 3pt}
\hline \hline
\end{tabular*}
\label{Table:Microscopic_symmetry_band_geometry}
\end{table}
%

%
\begingroup
\setlength{\tabcolsep}{4 pt}
\renewcommand{\arraystretch}{0.8}
\begin{table*}[t!]
\caption{Magnetic point-group symmetry restrictions on the $\mathcal{T}$-odd and $\mathcal{T}$-even orbital susceptibility tensors. A cross (\xmark) and a tick (\cmark) indicate that the corresponding response tensor is symmetry-forbidden and symmetry-allowed, respectively. Here, ${\cal M}_{a}$, ${\cal C}_n^a$, and ${\cal S}_n^a$ denote the mirror, $n$-fold rotation, and $n$-fold roto-reflection symmetry operations about the $a$ axis ($a=x,y,z$), respectively.}
\begin{tabular}{c c c c c c c c c c c c c c c c c c}
\hline \hline 
\noalign{\vskip 6pt}
$\chi^{\nu;a}$  & ${\cal M}_x\cal{T}$ & ${\cal M}_y\cal{T}$ & ${\cal M}_z\cal{T}$ & ${\cal C}^x_2\cal{T}$ & ${\cal C}^y_{2}\cal{T}$ & ${\cal C}^z_{2}\cal{T}$ & ${\cal C}^x_4\cal{T}$ & ${\cal C}_4^y\cal{T}$ & ${\cal C}_{3,6}^x\cal{T}$ & ${\cal C}_{3,6}^y\cal{T}$ & ${\cal C}_{3,4,6}^z\cal{T}$ & ${\cal S}_4^x\cal{T}$  & ${\cal S}_6^x\cal{T}$ & ${\cal S}_4^y\cal{T}$  & ${\cal S}_6^y\cal{T}$ & ${\cal S}_{4,6}^z\cal{T}$  \\
\noalign{\vskip 6pt}
\hline \hline 

\noalign{\vskip 6pt}

$\chi^{z,x}_{\mathrm{odd}}$ & \xmark  & \cmark & \xmark & \cmark & \xmark & \cmark & \xmark & \cmark & \xmark & \xmark & \xmark & \xmark & \xmark & \cmark & \cmark & \xmark  \\

\noalign{\vskip 6pt}

$\chi^{z,y}_{\mathrm{odd}}$  & \cmark & \xmark & \xmark & \xmark & \cmark & \cmark & \cmark & \xmark & \xmark & \xmark & \xmark & \cmark & \cmark & \xmark & \xmark & \xmark  \\

\noalign{\vskip 6pt}

\hline

\noalign{\vskip 6pt}

$\chi^{z,x}_{\mathrm{even}}$ & \cmark  & \xmark & \cmark & \xmark & \cmark & \xmark & \xmark & \cmark & \xmark & \cmark & \xmark & \xmark & \xmark & \cmark & \xmark & \xmark  \\

\noalign{\vskip 6pt}

$\chi^{z,y}_{\mathrm{even}}$  & \xmark & \cmark & \cmark & \cmark & \xmark & \xmark & \cmark & \xmark & \cmark & \xmark & \xmark & \cmark & \xmark & \xmark & \xmark & \xmark
\\

\noalign{\vskip 6pt} 
\hline \hline
\end{tabular}
\label{table_mag_point_group}
\end{table*}
\endgroup


In this section, we analyze the fundamental and crystallographic symmetry properties of the orbital susceptibility tensor. Under space inversion, the electric field transforms as $\bm E\rightarrow-\bm E$, whereas the orbital magnetization, being an axial vector, remains invariant, $\delta\bm L\rightarrow\delta\bm L$. Thus, the linear orbital susceptibility, $\delta L^{\nu}=\chi^{\nu,a}E_a$, is forbidden in a centrosymmetric system. To determine the behavior under time reversal, we examine the momentum-space parity of the susceptibility integrands in Eqs.~(\ref{chi_intrinsic}), (\ref{chi_sj}), (\ref{chi_sk3}), and (\ref{chi_sk4}) using the transformation properties of the underlying band-geometric quantities listed in Table~\ref{Table:Microscopic_symmetry_band_geometry}. We find that the intrinsic, side-jump, and skew-scattering contributions are $\cal T$-odd, whereas the conventional Drude contribution is $\cal T$-even. Consequently, in a noncentrosymmetric, nonmagnetic system, only the conventional Edelstein contribution is allowed. In an inversion-broken magnetic system, all five channels, intrinsic (anomalous), Drude, side-jump, third-order skew scattering, and fourth-order skew scattering, can be symmetry allowed, subject to the crystallographic point group.

In addition to the fundamental symmetry restrictions, the response is further constrained by the crystallographic point group of the material. From the constitutive relation, $\delta L^{\nu}=\chi^{\nu,a}E_a$, it follows that the response tensor $\chi^{\nu,a}$ is a second-rank axial tensor. Under magnetic point groups, the $\mathcal{T}$-even and $\mathcal{T}$-odd parts of this response tensor transform differently,
\bea
\chi^{\nu,a}_{\mathrm{even}} &=& \eta_R {\mathcal O}_{\nu \beta}{\mathcal O}_{a b}~\chi^{\beta,b}_{\mathrm{even}}~, \\
\chi^{\nu,a}_{\mathrm{odd}} &=& \eta_{\mathcal{T}} \eta_R {\mathcal O}_{\nu \beta}{\mathcal O}_{ab}~\chi^{\beta,b}_{\mathrm{odd}}~.
\eea
Here, $\eta_{\mathcal{T}}=+1$ for purely spatial operations ($\mathcal O=R$) and $\eta_{\mathcal{T}}=-1$ for symmetry operations involving time reversal ($\mathcal O=R\mathcal T$). The factor $\eta_R=\det{\mathcal O}$ accounts for the axial nature of the response tensor. In two-dimensional systems, the conventional OAM is oriented perpendicular to the plane, and accordingly, the only independent susceptibility components are $\chi^{z,x}$ and $\chi^{z,y}$. The resulting magnetic-point-group constraints on the $\mathcal{T}$-even conventional contribution and on the $\mathcal{T}$-odd intrinsic, side-jump, and skew-scattering contributions are summarized in Table~\ref{table_mag_point_group}.


\section{OEE in a magnetized Rashba 2DEG} \label{sec:Rashba_2DEG}


To demonstrate the material relevance of our theory and quantify the EOEE, we consider a 2DEG with Rashba SOC, which can be realized at various $\ch{SrTiO3}$-based interfaces, including $\ch{LaAlO3/SrTiO3}$~\cite{Triscone_10_prl, Han_17_sciadv, Johansson_2021}, $\ch{LaTiO3/SrTiO3}$~\cite{Ohtomo_02_nature, Rotenberg_13_prl}, and $\ch{BaTiO3/SrTiO3}$~\cite{Ping_16_prb, Demkov_15_prb}. In a nonmagnetic system, only the $\cal T$-even conventional Drude contribution to the OEE is symmetry allowed, while the intrinsic and asymmetric-scattering contributions vanish. We therefore introduce an exchange field to break $\cal T$ symmetry and open a gap. The minimal Hamiltonian is~\cite{kohda_2019_sreport, Ishizaka_2011_ncom, SARKAR_2026_MTQ, Rahul_2026}
\begin{align}
    \mathcal{H} = \frac{\hbar^2 k^2}{2m}+ \alpha \left(k_y \sigma_x - k_x \sigma_y\right)
    + \bm{M}\cdot\bm{\sigma}~.\label{Rashba_2DEG_Ham}
\end{align}
Here, $\alpha$ is the Rashba SOC strength, $m$ is the effective mass, and $\sigma^\nu$ ($\nu=x,y,z$) denotes a Pauli matrix acting on the spin degree of freedom. The last term describes the exchange coupling, with $\bm M$ denoting the exchange field.

Although broken inversion and $\cal T$ symmetries are necessary for the $\cal T$-odd OEE, they do not by themselves guarantee a finite response. The combined ${\cal C}_n^z{\cal T}$ symmetries with $n=3,4,6$ forbid both $\cal T-$odd and $\cal T-$even components of the response tensor $\chi^{z,x}$ and $\chi^{z,y}$ [see Table~\ref{table_mag_point_group}]. A nonzero in-plane component of $\bm M$ is therefore required. An in-plane exchange field alone preserves ${\cal C}_2^z{\cal T}$, which forces the band-diagonal out-of-plane OAM and Berry curvature (BC) to vanish. This constraint removes the response channels constructed from these quantities but does not generally eliminate interband contributions such as the anomalous response. We retain both components of $\bm M$ to maintain a finite gap and finite band-diagonal geometric quantities. We accordingly consider the exchange field $\bm{M} = M(\sin\theta,\,0,\,\cos\theta)$, where $\theta$ is the angle between the $\bm M$ and the $z$ axis. This configuration preserves the combined ${\cal M}_y {\cal T}$ symmetry, which allows only the $\chi^{z,x}_{\rm odd}$ and $\chi^{z,y}_{\rm even}$ components of the OEE tensor, as listed in Table~\ref{table_mag_point_group}. Since we focus on the $\mathcal T$-odd extrinsic response, we consider the electric field along $x$ and discuss the $\chi^{z,x}_{\rm odd}$ component below.

To obtain the band dispersion and the $\bm{k}$-space geometric quantities, it is convenient to rewrite the Hamiltonian (\ref{Rashba_2DEG_Ham}) in the compact form
\begin{align}
    \mathcal{H} = \varepsilon_{\bm k} + \bm{d}\cdot\bm{\sigma}~,\label{Ham_d_vec}
\end{align}
where $\varepsilon_{\bm k}=\hbar^2k^2/2m$ is the free-electron kinetic energy and the components of the $\bm{d}$ vector are $ d_x = \alpha k_y + M\sin\theta,
    ~d_y = -\alpha k_x,~ 
    d_z = M\cos\theta.$
Within this notation, the energy dispersion of the two Rashba-split bands is given by
\begin{align}
    \varepsilon_{n\bm k}=\varepsilon_{\bm k} + n d(\bm k)~,
\end{align}
where $d(\bm k) = \sqrt{d_x^2+d_y^2+d_z^2}
$ and $n = 1$ ($-1$) labels the conduction (valence) band. For the Hamiltonian of the form in Eq.~\eqref{Ham_d_vec}, the OAM, BC, and energy-normalized BC are given by
\begin{align}
    L^z_{n\bm k} &= \frac{-e}{2 \mu_B} \frac{\bm d \cdot (\partial_x \bm d \times \partial_y \bm d)}{d^2} =-  \frac{e}{2 \mu_B} \frac{\alpha^2 M \cos\theta}{d^2}~, \label{L_z_2DEG}
\end{align}
\begin{align}
    \Omega^z_{n\bm k} &=- \frac{n}{2} \frac{\bm d \cdot (\partial_x \bm d \times \partial_y \bm d)}{d^3} = - \frac{n}{2} \frac{\alpha^2 M \cos\theta}{d^3}~, \label{BC_z_2DEG}
\end{align}
\begin{align}
    \tilde{\Omega}^z_{n\bm k} &=- \frac{1}{4} \frac{\bm d \cdot (\partial_x \bm d \times \partial_y \bm d)}{d^4} = - \frac{1}{4} \frac{\alpha^2 M \cos\theta}{d^4}~. \label{BC_normalized_z_2DEG}
\end{align}
\begin{figure}[b]
    \centering
    \includegraphics[width=\linewidth]{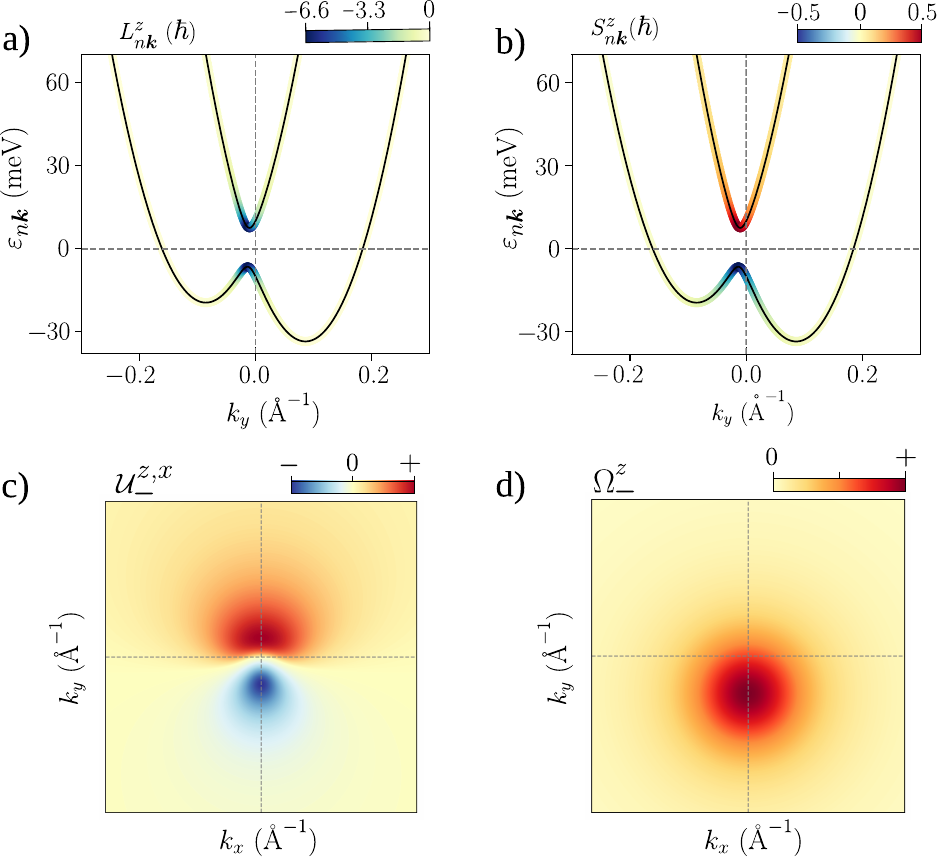}
    \caption{\textbf{Band dispersion and geometric quantities of the Rashba 2DEG.} (a) and (b) show the $z$-component of the orbital- and spin-angular-momentum-projected band dispersions along the $k_y$ direction, respectively. (c) and (d) present the momentum-space distributions of the AOP and BC, respectively, for the outer band ($n=-1$). The calculations are performed using the parameter set $ \alpha = 0.6~\mathrm{eV\AA}$, $M = 10~\mathrm{meV}$, $\theta = \pi/4$, and an effective mass $m = 1.1m_e$, where $m_e$ is the bare electron mass.}
    \label{fig:Fig_2}
\end{figure}
\begin{figure}
    \centering
    \includegraphics[width=\linewidth]{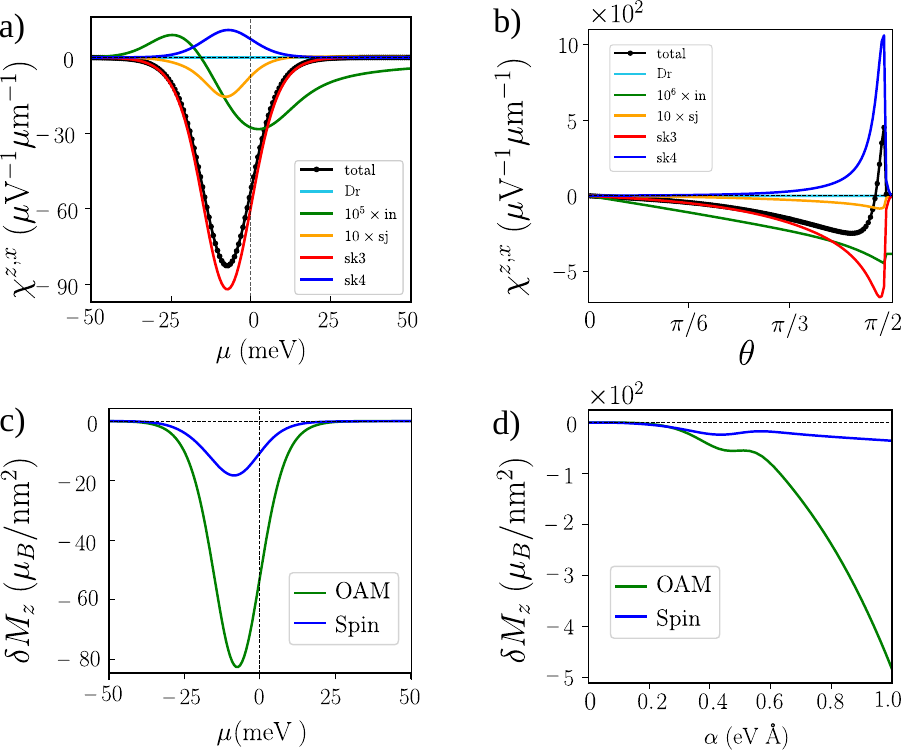}
    \caption{\textbf{Variation of the response with model parameters.} (a) and (b) show the dependence of the contributions to $\chi^{z,x}$ on the chemical potential and exchange-field orientation $\theta$, respectively. For better visibility, the side-jump contribution is multiplied by \(10\), while the intrinsic contribution is multiplied by \(10^5\) and \(10^6\) in panels (a) and (b), respectively. (c) and (d) show the total orbital and spin magnetizations obtained from the five different contributions as functions of chemical potential and Rashba SOC strength, respectively. For (a) and (c), the parameters are the same as in Fig.~\ref{fig:Fig_2}. For (b), $\alpha = 0.6~\mathrm{eV\,\AA}$, $M = 10~\mathrm{meV}$, and $\mu = -7.5~\mathrm{meV}$. For (d), $M = 10~\mathrm{meV}$, $\theta = \pi/4$, and $\mu = -7.5~\mathrm{meV}$. The induced magnetization is calculated for $\bm{E}=10^6~\mathrm{V/m}\,\hat{\bm{x}}$.} 
    \label{fig:Fig_3}
\end{figure}
These expressions show that the band-geometric quantities are enhanced near the avoided crossing, where the band gap is minimal. As shown in Fig.~\ref{fig:Fig_2}(a), the exchange field shifts the avoided crossing from $(0,0)$ to $(0,-M\sin\theta/\alpha)$, highlighting the breaking of in-plane rotational symmetry by the in-plane component of $\bm M$. The OAM reaches $\sim7\hbar$ near the band edges, whereas the spin angular momentum (SAM) in Fig.~\ref{fig:Fig_2}(b) remains bounded by $\pm\hbar/2$. The AOP and BC distributions in Figs.~\ref{fig:Fig_2}(c) and \ref{fig:Fig_2}(d) for the valence band exhibit a similar concentration near the avoided crossing and reflect the broken rotational symmetry. Owing to the $\mathcal{M}_y \cal T$ symmetry of the model, these quantities satisfy $\mathcal{U}^{z,x}_{n\bm k}(k_x,k_y)=\mathcal{U}^{z,x}_{n\bm k}(-k_x,k_y),$ and $\Omega^z_{n\bm k}(k_x,k_y)
= \Omega^z_{n\bm k}(-k_x,k_y),$ as is evident from the corresponding plots.

The OEE responses are shown in Fig.~\ref{fig:Fig_3}. Figure~\ref{fig:Fig_3}(a) shows the chemical-potential dependence of the different contributions to the response tensor. Owing to the $\mathcal{M}_y\mathcal{T}$ symmetry, the conventional Edelstein contribution, $\chi^{z,x}_{\rm Dr}$, vanishes identically. Because the AOP has opposite signs in the valence and conduction bands, $\mathcal{U}^{z,x}_{-}=-\mathcal{U}^{z,x}_{+}$, the intrinsic response reverses sign as the chemical potential crosses the avoided crossing. Away from the avoided crossing, the contributions from the two bands largely cancel, suppressing the intrinsic response. For the numerical evaluation of the side-jump and skew-scattering contributions, we take an impurity density of $n_i=10^{10}~\mathrm{cm}^{-2}$ and an impurity potential strength of $V_0=6.2\times10^{-13}~\mathrm{eV, cm}^{2}$, corresponding to a symmetric scattering time of approximately $\tau\simeq0.1~\mathrm{ps}$. To compare the third- and fourth-order skew-scattering contributions, we set the first-order moment of the impurity potential to $V_1=0.15V_0$. Similar parameter values have been used in Refs.~\cite{Justin_2023_prl, Fu_2021_prb, Harsh_2026_prb}. Since the side-jump and skew-scattering terms are Fermi-surface contributions [Eqs.~(\ref{chi_sj})--(\ref{chi_sk4})], they are strongly enhanced when the chemical potential lies near the avoided crossing, where the OAM reaches its maximum magnitude. As the chemical potential moves away from this region, the OAM rapidly decreases, leading to a corresponding suppression of both contributions.

\subsection{Exchange-field tunability of the EOEE}

Figure~\ref{fig:Fig_3}(b) shows the dependence of the OEE responses on the orientation of the exchange field. When the exchange field is aligned along the $z$ axis, all OEE response components vanish. This behavior is consistent with the symmetry analysis in Table~\ref{table_mag_point_group}, since for $\bm M\parallel\hat{\bm z}$, the system preserves both ${\cal M}_x\cal{T}$ and ${\cal M}_y\cal{T}$ symmetries, which forbid the OEE responses.
In the other limiting case, $\bm M\perp\hat{\bm z}$, the system preserves the ${\cal C}_2^z{\cal T}$ symmetry, which forces the band-diagonal OAM, BC, and energy-normalized BC to vanish throughout momentum space. Consequently, the conventional and asymmetric-scattering contributions constructed from these quantities vanish. The anomalous contribution can remain finite because it is governed by interband matrix elements of the OAM operator rather than the band-diagonal OAM.

For $0<\theta<\pi/2$, the response is finite due to the breaking of ${\cal C}_n^z{\cal T}$ $(n=2,3,4,6)$ symmetries and increases as the exchange field approaches the plane of the 2DEG. This enhancement originates from the decreasing gap $\Delta{\rm gap}=2M\cos\theta$, which strongly amplifies the associated interband geometric quantities [see Eqs.~(\ref{L_z_2DEG})--(\ref{BC_normalized_z_2DEG})] and, consequently, the disorder-induced contributions to the OEE.



\subsection{Orbital versus spin Edelstein effects and their tuning with Rashba coupling}


Finally, we compare the OEE-induced magnetization with its spin counterpart. The total magnetization decomposes into orbital and spin contributions, $\delta M^{\nu} = \delta L^{\nu} + \delta S^{\nu}$, where the orbital part is given by Eq.~\eqref{eq: def_OEE} and the spin part is~\cite{Cong_Xiao_24_prb}
\begin{align}
    \delta S^{\nu} = - \frac{g_s \mu_B}{\hbar} \sum_l \tilde{s}^{\nu}_l f_l~.
\end{align}
Here, $\tilde{s}^{\nu} = (\hbar/2) \braket{\tilde{l}| \sigma^{\nu}|\tilde{l}}$ is the SAM carried by a perturbed Bloch state and $g_s \simeq 2$ for electrons. The spin Edelstein effect follows from the same procedure used for the OAM. Figure \ref{fig:Fig_3}(c) shows the chemical-potential dependence of the orbital and spin contributions to the total
magnetization, and Fig. \ref{fig:Fig_3}(d) shows their dependence on the Rashba coupling strength. The two contributions are comparable in the weak-SOC regime, whereas the orbital contribution increasingly dominates as $\alpha$ increases. The origin of this asymmetry is that the SAM is bounded by $\pm \hbar/2$, so the spin Edelstein effect saturates once the spin texture is fully polarized, while the OAM carries no such bound: increasing $\alpha$ enhances the interband velocity matrix elements, and the OAM at the avoided crossing grows as $L^z_{\rm max} \propto \alpha^2/(M \cos\theta)$. Numerically, we find that $\delta L^z$ grows approximately quadratically with $\alpha$ in the strong-SOC regime. For the adopted disorder parameters and a realistic Rashba strength~\cite{Ishizaka_2011_ncom, Eremeev_2012_prl, Wanjun_2022_nanolett} $\alpha = 1~\mathrm{eV\,\AA}$, the orbital contribution is almost $13$ times larger than the spin contribution. This highlights the dominant role of orbital degrees of freedom in current-induced magnetization at realistic Rashba SOC strengths.


\section{Conclusion} \label{sec:conclusion}


In summary, we have developed a microscopic semiclassical Boltzmann theory of the orbital Edelstein effect that goes beyond the conventional relaxation-time approximation and treats the intrinsic band-geometric and disorder-induced mechanisms on equal footing. We identify two extrinsic channels: the side-jump contribution arising from the field-induced coordinate shift and the skew-scattering contribution generated by antisymmetric impurity scattering. The resulting orbital susceptibility naturally decomposes into conventional Drude, intrinsic, side-jump, and skew-scattering contributions, with distinct scattering-time dependences. Our symmetry analysis further establishes the symmetry requirements for realizing the different OEE contributions. As a concrete realization, we considered a magnetized Rashba two-dimensional electron gas and demonstrated that the interplay of Rashba spin--orbit coupling and the exchange field orientation produces large orbital angular momentum near the avoided band crossing. Within the adopted disorder approximation and for the selected parameters, the corresponding band-geometric enhancement strongly amplifies the side-jump and skew-scattering contributions, which can substantially exceed the intrinsic and conventional responses. Moreover, the orbital response increases strongly with Rashba spin--orbit coupling and, for realistic SOC strengths, exceeds the spin Edelstein contribution by approximately one order of magnitude, reaching nearly a factor of 13 for \( \alpha = 1 ~\text{eV} \mathrm{\AA}\). Our results show that disorder scattering can substantially modify the orbital Edelstein effect and provide a route to controlling current-induced orbital magnetization.

\section*{Acknowledgments}

We acknowledge many fruitful discussions with Harsh Varshney (IIT Kanpur, India) and
Sayan Sarkar (IIT Kanpur, India). S.S. acknowledges financial support from the Indian Institute of Technology Kanpur. K.G. is supported by the Ministry of Education, Government of India, through the Prime Minister's Research Fellowship. A.A. acknowledges funding from the Core Research Grant by ANRF (Sanction No. CRG/2023/007003), Department of Science and Technology, India.


\begin{appendix}


\onecolumngrid

\section{Side-jump contribution to OAM}\label{side_jump_L}
Following the discussion in the main text, the presence of the impurity potential modifies the Bloch states of the unperturbed Hamiltonian, $\mathcal{H}$. In the absence of an external electric field, the eigenstate of the total Hamiltonian, $H = \mathcal{H} + V_{\rm imp},$ can be written as $\ket{l_{\rm dis}} = \ket{l} + \ket{\delta l_{\rm dis}},$ where $\ket{l}$ is the eigenstate of the unperturbed Hamiltonian and $\ket{\delta l_{\rm dis}}$ denotes the correction induced by the impurity potential $V_{\rm imp}$. The perturbed eigenstate $\ket{l_{\rm dis}}$ is obtained from the Lippmann--Schwinger equation \cite{sakurai2020modern},
\begin{equation}
    \ket{l_{\rm dis}} = \ket{l} + (\varepsilon_l - \mathcal{H} + i \eta) ^{-1} V_{\rm imp} \ket{l_{\rm dis}}~,
\end{equation}
where the infinitesimal $i \eta \to 0^{+}$ ensures the outgoing
boundary condition for the scattering states. The side-jump correction to the OAM for the state $\ket{l}$ is given by \cite{Cong_Xiao_24_prb, sanjay_2026_prb},
\begin{align}
    L^{\nu, \rm sj}_l = \Big\langle 2 \text{Re} \bra{l} \hat{L}^{\nu} \ket{\delta l_{\rm dis}} + \bra{\delta l_{\rm dis}} \hat{L}^{\nu} \ket{\delta l_{\rm dis}} \Big\rangle_{\rm dis}~.
\end{align}
The disorder average, $\langle \cdots \rangle_{\rm dis}$, over impurity configurations is essential for restoring translational invariance, allowing the theory to be formulated in the crystal momentum representation. 

For randomly distributed impurity potentials, the leading nonzero contribution to the side-jump response originates from terms that are second order in the disorder potential, giving
\begin{align}
        L^{\nu, \rm sj}_l &= 2 \text{Re} \sum_{l', l''} \frac{\braket{l | \hat{L}^{\nu} | l'} \langle V_{l' l''} V_{l'' l} \rangle_{\rm dis}}{(\varepsilon_l - \varepsilon_{l'} + i \eta) ( \varepsilon_l - \varepsilon_{l''} + i \eta )}  + \sum_{l', l''} \frac{ \braket{l' | \hat{L}^{\nu} | l''} \langle V_{l l'} V_{l'' l} \rangle_{\rm dis} }{(\varepsilon_l - \varepsilon_{l'} - i \eta) ( \varepsilon_l - \varepsilon_{l''} + i \eta )}~.
\end{align}
This expression follows by iterating the Lippmann--Schwinger equation to second order in $V_{\rm imp}$. We introduce the notation $V_{ll'}= \braket{l|V_{\rm imp}|l'}= \braket{n\bm{k}|V_{\rm imp}|n'\bm{k}'}$.
To proceed further, we specify the nature and configuration of the disorder potential. In this work, we model the disorder as a collection of randomly distributed short-range \(\delta\)-function impurities, described by
\begin{align}
    V_{\rm imp}(\bm{r})=\sum_i V_i\,\delta(\bm{r}-\bm{R}_i)~.
\end{align}
Here, $V_i$ denotes the potential strength at site $\bm R_i$. The delta-function model fixes the disorder cumulants. Below, an intraband small-$\bm q$ expansion of the Bloch-state overlaps selects the local scattering sector. The corresponding matrix element is
\begin{align}
    V_{l l'} &= \sum_i \int d \bm r ~ V_i \delta(\bm r - \bm R_i) e^{i (\bm k' - \bm k) \cdot \bm r} \braket{u_{n \bm k} | u_{n' \bm k'}} \nn \\
    &= \sum_i V_i e^{i (\bm k' - \bm k) \cdot \bm R_i} \braket{u_{n \bm k}|u_{n' \bm k'}} = V^0_{\bm k \bm k'} u^{\bm k \bm k'}_{n n'}~,
\end{align}
where we have defined $ u^{\bm k \bm k'}_{n n'} = \braket{u_{n \bm k} | u_{n' \bm k'}} $. Using the momentum-space representation of the OAM operator, the matrix element of $\hat{L}^{\nu}$ in the Bloch state can be written as $\braket{l | \hat{L}^{\nu} | l'} = \braket{n \bm k | \hat{L}^{\nu} | n' \bm k'} = L^{\nu}_{ n n'} (\bm k) \delta(\bm k' - \bm k)~.$ Inserting these simplifications into the expression for $L^{\nu, \rm sj}_l$ gives
\begin{align}
    L^{\nu, \rm sj}_{n \bm k} &= 2 \text{Re} \sum_{n', n'', \bm k', \bm k''} \frac{L   ^{\nu}_{n n'} (\bm k) \delta(\bm k - \bm k') \langle V^0_{\bm k \bm k''} V^0_{\bm k'' \bm k} \rangle_{\rm dis} u^{\bm k \bm k''}_{n' n''} u^{\bm k'' \bm k}_{n'' n} }{(\varepsilon_{n \bm k} - \varepsilon_{n' \bm k}) (\varepsilon_{n \bm k} - \varepsilon_{n'' \bm k''} + i \eta)} + \sum_{n', n'', \bm k', \bm k''} \frac{L^{\nu}_{n' n''} (\bm k'') \delta(\bm k' - \bm k'') \langle V^0_{\bm k \bm k''} V^0_{\bm k'' \bm k} \rangle_{\rm dis} u^{\bm k \bm k''}_{n n'} u^{\bm k'' \bm k}_{n'' n} }{(\varepsilon_{n \bm k} - \varepsilon_{n' \bm k''} - i \eta) (\varepsilon_{n \bm k} - \varepsilon_{n'' \bm k''} + i \eta)} \nn \\
    &= 2 \text{Re} \sum_{n', n'', \bm k'}  \frac{L^{\nu}_{n n'} (\bm k) \langle V^0_{\bm k \bm k'} V^0_{\bm k' \bm k} \rangle_{\rm dis}  u^{\bm k \bm k'}_{n' n''} u^{\bm k' \bm k}_{n'' n} }{(\varepsilon_{n \bm k} - \varepsilon_{n' \bm k}) (\varepsilon_{n \bm k} - \varepsilon_{n'' \bm k'} + i \eta)} + \sum_{n', n'', \bm k'} \frac{L^{\nu}_{n' n''} (\bm k')  u^{\bm k \bm k'}_{n n'} u^{\bm k' \bm k}_{n'' n} }{(\varepsilon_{n'' \bm k'} - \varepsilon_{n' \bm k'})} \Big( \frac{\langle V^0_{\bm k \bm k'} V^0_{\bm k' \bm k} \rangle_{\rm dis}}{(\varepsilon_{n \bm k} - \varepsilon_{n'' \bm k'} + i \eta)} - \frac{\langle V^0_{\bm k \bm k'} V^0_{\bm k' \bm k} \rangle_{\rm dis}}{(\varepsilon_{n \bm k} - \varepsilon_{n' \bm k'} - i \eta)} \Big)  \nn \\
    &= 2 \text{Re} \sum_{n', n'', \bm k'} \Big[ \frac{L^{\nu}_{n n'} (\bm k) \langle V^0_{\bm k \bm k'} V^0_{\bm k' \bm k} \rangle_{\rm dis}  u^{\bm k \bm k'}_{n' n''} u^{\bm k' \bm k}_{n'' n} }{(\varepsilon_{n \bm k} - \varepsilon_{n' \bm k}) (\varepsilon_{n \bm k} - \varepsilon_{n'' \bm k'} + i \eta)} + \frac{L^{\nu}_{n'' n'} (\bm k') \langle V^0_{\bm k \bm k'} V^0_{\bm k' \bm k} \rangle_{\rm dis}  u^{\bm k \bm k'}_{n n''} u^{\bm k' \bm k}_{n' n} }{(\varepsilon_{n'' \bm k'} - \varepsilon_{n' \bm k'}) (\varepsilon_{n \bm k} - \varepsilon_{n'' \bm k'} - i \eta)} \Big] \nn \\
    &= - 2 \pi \sum_{n', \bm k'} \left\langle V^{0}_{\bm{k}\bm{k}'}V^{0}_{\bm{k}'\bm{k}}\right\rangle_{\rm dis} \delta( \varepsilon_{n \bm k} - \varepsilon_{n' \bm k'} )~ \text{Im} \Bigg[ \sum_{n'' \neq n'} \frac{L^{\nu}_{n' n''} (\bm k')   u^{\bm k \bm k'}_{n n'} u^{\bm k' \bm k}_{n'' n} }{(\varepsilon_{n' \bm k'} - \varepsilon_{n'' \bm k'})} - \sum_{n'' \neq n} \frac{L^{\nu}_{n n''} (\bm k)  u^{\bm k \bm k'}_{n'' n'} u^{\bm k' \bm k}_{n' n} }{(\varepsilon_{n \bm k} - \varepsilon_{n'' \bm k}) } \Bigg]~.
\end{align}
In the last step, we use $\lim_{\eta \to 0} \operatorname{Im}~(x \pm i\eta)^{-1}= \mp \pi \delta(x)$. It is interesting to note that the side-jump contribution is closely related to the AOP,
$\mathcal{U}^{\nu,a}_{n\bm{k}}$. To establish the connection, we use the intraband small-momentum-transfer approximation,
$\bm{q}=\bm{k}'-\bm{k}\rightarrow 0$ and scattering is predominantly intraband ($n' = n$).
In this case, the overlaps of the cell-periodic parts of the Bloch states can be expanded in powers of $\bm q$. To linear order, this gives
\begin{align}
    u^{\bm k \bm k'}_{n n} = 1 - i q_a \mathcal{R}^a_{nn}~,~ u^{\bm k \bm k'}_{n n''} = - i q_a \mathcal{R}^a_{n n''}~, ~ u^{\bm k \bm k'}_{n '' n} = - i q_a \mathcal{R}^a_{n'' n}~. 
\end{align}
Here, $ \bm{\mathcal{R}}_{nn'} = i \braket{u_{n\bm k}| \partial_{\bm k} | u_{n'\bm k}} $ is the Berry connection. Substituting these overlaps into the preceding expression for
$L^{\nu,\rm sj}_{n\bm k}$
and retaining terms linear in $\bm q$ gives
\begin{align}
    L^{\nu, \rm sj}_{n \bm k} &= - 2 \pi \sum_{\bm k'} q_a  \left\langle V^{0}_{\bm{k}\bm{k}'}V^{0}_{\bm{k}'\bm{k}}\right\rangle_{\rm dis} \delta( \varepsilon_{n \bm k} - \varepsilon_{n \bm k'} ) \times \text{Im} \sum_{n'' \neq n} \Big[ \frac{i L^{\nu}_{n n''} (\bm k') \mathcal{R}^a_{n'' n} (\bm k')}{(\varepsilon_{n \bm k'} - \varepsilon_{n'' \bm k'})} + \frac{i L^{\nu}_{n n''} (\bm k) \mathcal{R}^a_{n'' n} (\bm k)}{(\varepsilon_{n \bm k} - \varepsilon_{n'' \bm k})} \Big] \nn \\
    &=  \frac{2 \pi}{\hbar} \sum_{\bm k'}  \left\langle V^{0}_{\bm{k}\bm{k}'}V^{0}_{\bm{k}'\bm{k}}\right\rangle_{\rm dis} \delta( \varepsilon_{n \bm k} - \varepsilon_{n \bm k'} )[ (k_a - k_a') ~\mathcal{U}^{\nu, a}_{n \bm k}]~.
\end{align}
The second line follows from  $\mathcal R^a_{nn''}(\bm k)=-i\hbar v^a_{nn''}(\bm k)/[\varepsilon_{n\bm k}-\varepsilon_{n''\bm k}]$ and the definition of $\mathcal U^{\nu,a}_{n\bm k}$.
Returning to the disorder model introduced above, the disorder average $\left\langle V^{0}_{\bm{k}\bm{k}'}V^{0}_{\bm{k}'\bm{k}}\right\rangle_{\rm dis}$ reduces to
\begin{align}
    W_{\bm k\bm k'} =  \left \langle 
V^0_{\kb\kb'} V^0_{\kb'\kb}\right \rangle_{\text{dis}}=\left\langle\sum_{ij}V_i V_j \,\exp[i(\kb'-\kb)\cdot (\bm R_i-\bm R_j)]\right\rangle_{\rm dis}=n_iV_0^2~.
\end{align}
Here, $n_i$ is the impurity density, and $V_0$ represents the zeroth-order moment of the impurity potential, with units of energy times area for the two-dimensional disorder model.
%
%
%
\section{Non-equilibrium distribution function}\label{non_equilibrium_distributiom}
%
%
In this section, we derive the nonequilibrium Fermi--Dirac distribution function to linear order in the external electric field, $\bm{E}$, including elastic impurity scattering. The distribution function $f_l$ satisfies
\begin{equation}
    \frac{\partial f_l}{\partial t} + \dot{\bm k} \cdot \partial_{\bm k} f_l = I_{\rm el} \{ f_l \}~,
\end{equation}
where $I_{\mathrm{el}}\{f_l\} = -\sum_{l'} \left( w_{l'l}f_l - w_{ll'}f_{l'}\right)$ is the elastic collision integral and $w_{ll'}$ denotes the scattering rate for $l'\to l$. In the presence of a dc electric field, the steady-state condition implies $\partial f_l/\partial t = 0$, reducing the Boltzmann equation to $\dot{\bm{k}} \cdot \partial_{\bm{k}} f_l = I_{\rm el}\{f_l\}.$ As discussed in the main text, the scattering rates $w_{ll'}$ are generally not symmetric under the exchange of the initial and final states. Consequently, they can be decomposed into symmetric and antisymmetric parts as
\begin{equation}
    w_{ll'}^{\rm S} = w_{l'l}^{\rm S} =  \frac{w_{ll'} + w_{l'l}}{2}~, ~~w_{ll'}^{\rm A} = - w_{l'l}^{\rm A} = \frac{w_{ll'} - w_{l'l}}{2}~.
\end{equation}
The symmetric component, $w_{ll'}^{\rm S}$, governs the conventional relaxation of the nonequilibrium distribution and is typically treated within the relaxation-time approximation. In contrast, the antisymmetric component, $w_{ll'}^{\rm A}$, is responsible for the skew-scattering contribution to transport. 

To determine the nonequilibrium distribution function in the presence of different scattering processes, we consider the weak-disorder limit and expand the scattering rates in powers of the impurity potential strength as
\begin{align}
    w_{ll'} = w_{ll'}^{(2)} + w_{ll'}^{(3),\rm A} + w_{ll'}^{(4),\rm A}~.
\end{align}
Here, $w_{ll'}^{(3),\rm A}$ and $w_{ll'}^{(4),\rm A}$ denote the antisymmetric scattering rates arising at third and fourth order in the impurity potential, respectively. The second-order scattering rate can be further decomposed into a field-independent symmetric contribution ($w_{l'l}^{(2),\rm S}$) and an electric-field-induced coordinate-shift contribution ($w_{l'l}^{(2), \rm cs}$),
\begin{align}
    w_{l'l}^{(2),\rm S} &= \frac{2 \pi}{\hbar} \langle | V_{l'l} |^2 \rangle_{\rm dis} \delta( \varepsilon_l - \varepsilon_{l'} )~, \\
    w_{l'l}^{(2), \rm cs} &=- \frac{2 \pi}{\hbar} \langle |V_{l'l}|^2 \rangle_{\rm dis} \frac{ \partial \delta( \varepsilon_l - \varepsilon_{l'} )}{\partial \varepsilon_{l}} e\bm E \cdot \delta \bm r_{l'l}~.
\end{align}
Using these distinct scattering rates, the collision integral can be decomposed as
$ I_{\rm el}\{f_l\} = I_{\rm el}^{\rm Dr} \{f_l \} + I_{\rm el}^{\rm sj} \{f_l \} + I_{\rm el}^{\rm sk3} \{f_l \} + I_{\rm el}^{\rm sk4} \{f_l \} $. Their explicit expressions are given by
\begin{align}
    &I_{\rm el}^{\rm Dr} \{f_l \} = - \sum_{l'} w_{ll'}^{(2), \rm S} (f_l - f_{l'})~, \\
    &I_{\rm el}^{\rm sj} \{f_l \} =  - \sum_{l'} w_{ll'}^{(2), \rm cs} (f_l - f_{l'})~, \\
    &I_{\rm el}^{\rm sk3} \{f_l \} =   \sum_{l'} w_{ll'}^{(3), \rm A} (f_l + f_{l'})~, \\
    &I_{\rm el}^{\rm sk4} \{f_l \} =   \sum_{l'} w_{ll'}^{(4), \rm A} (f_l + f_{l'})~. 
\end{align}
We decompose the distribution in the same way, $f_l=f_l^{\rm Dr}+f_l^{\rm sj}+f_l^{\rm sk3}+f_l^{\rm sk4}$, where $f_l^{\rm Dr}=f_l^0+f_l^{\rm Dr,(1)}+\mathcal O(E^2)$ and the other components start at linear order. For a spatially uniform electric field, $\dot{\bm k}=-e\bm E/\hbar$, and the four coupled equations for these components are

\begin{subequations}\label{BZ_coupl_eq:2}
\begin{align}
-\frac{e\bm E}{\hbar}\!\cdot\!\partial_{\bm k} f_l^{\rm Dr}
&= I_{\rm el}^{\rm Dr}(f_l^{\rm Dr})\label{BZ_eq1}~, \\
-\frac{e\bm E}{\hbar}\!\cdot\!\partial_{\bm k} f_l^{\rm sj}
&= I_{\rm el}^{\rm Dr}(f_l^{\rm sj}) + I_{\rm el}^{\rm sj}(f_l^{\rm Dr})\label{BZ_eq2}~, \\
-\frac{e\bm E}{\hbar}\!\cdot\!\partial_{\bm k} f_l^{\rm sk3}
&= I_{\rm el}^{\rm Dr}(f_l^{\rm sk3}) + I_{\rm el}^{\rm sk3}(f_l^{\rm Dr})\label{BZ_eq3}~, \\
-\frac{e\bm E}{\hbar}\!\cdot\!\partial_{\bm k} f_l^{\rm sk4}
&= I_{\rm el}^{\rm Dr}(f_l^{\rm sk4}) + I_{\rm el}^{\rm sk4}(f_l^{\rm Dr})\label{BZ_eq4}~.
\end{align}
\end{subequations}

We treat the symmetric second-order elastic collision integral within the relaxation-time approximation by introducing a characteristic relaxation time $\tau$, such that
\begin{equation}
I_{\rm el}^{\rm Dr}(f_l^{\rm Dr})
=
-\frac{f_l^{\rm Dr} - f_l^0}{\tau}~,
\end{equation}
where $f_l^0$ is the equilibrium Fermi--Dirac distribution function. To obtain the nonequilibrium distribution function perturbatively, we expand it in powers of the applied electric field,
\begin{equation}
f_l = f_l^{(0)} + f_l^{(1)} + f_l^{(2)} + \cdots~,
\end{equation}
where $f_l^{(0)} \equiv f_l^0$ and $f_l^{(i)} \propto |\bm{E}|^i$ $(i>0)$ denotes the $i$th-order correction to the distribution function. The conventional Drude linear-response solution is then readily obtained as
\begin{equation}
f_l^{\rm Dr,(1)}
=
\frac{e\tau}{\hbar}\,
\bm E\!\cdot\!\partial_{\bm k} f_l^0~.
\end{equation}

For the remaining channel corrections, the scalar relaxation-time approximation gives $I_{\rm el}^{\rm Dr}(f_l^{x,(1)})=-f_l^{x,(1)}/\tau$, with $x=\mathrm{sj},\mathrm{sk3},\mathrm{sk4}$. Solving Eq.~\eqref{BZ_eq2} to linear order in $\bm E$ yields the side-jump
correction
\begin{align}
f_l^{\rm sj,(1)}
&= -
\tau \sum_{l'} w^{(2),{\rm cs}}_{l'l}
\bigl(f_l^0 - f_{l'}^0\bigr) \nn \\
&= \tau \frac{2 \pi}{\hbar} \sum_{l'} \langle |V_{l'l}|^2\rangle_{\rm dis} \frac{\partial \delta(\varepsilon_l - \varepsilon_{l'})}{\partial \varepsilon_l} e\bm E \cdot \delta \bm r_{l'l}~(f_l^0 - f_{l'}^0)~.
\end{align}
Because $\partial_{\varepsilon_l}\delta(\varepsilon_l-\varepsilon_{l'})$ samples only the neighborhood of $\varepsilon_l=\varepsilon_{l'}$, only the linear term in the energy difference contributes:
\begin{align}
    f_l^0-f_{l'}^0
    =
    (\varepsilon_l-\varepsilon_{l'})
    \frac{\partial f_l^0}{\partial \varepsilon_l}~.
\end{align}
Using this linear term together with the identity
\begin{align}
    (\varepsilon_l-\varepsilon_{l'})
    \frac{\partial \delta(\varepsilon_l-\varepsilon_{l'})}
    {\partial \varepsilon_l}
    =
    -\delta(\varepsilon_l-\varepsilon_{l'})~,
\end{align}
we obtain
\begin{equation}
f_l^{\rm sj,(1)} = - e\tau\, \bm E\!\cdot\!\bm v_l^{\rm sj}\,\frac{\partial f^0}{\partial \varepsilon_l}~,
\end{equation}
where the side-jump velocity is
\begin{equation}\label{sj_vel_app}
\bm v_l^{\rm sj} = \sum_{l'} w^{(2),\rm S}_{l'l}\, \boldsymbol{\delta r}_{l'l}~.
\end{equation}
Following the same procedure, the third- and fourth-order antisymmetric scattering processes yield
\begin{align}
f_l^{\rm sk3,(1)} &= \frac{e\tau^2}{\hbar} \sum_{l'} w^{(3),\rm A}_{ll'} \left(
\bm E\!\cdot\!\partial_{\bm k} f_l^0 + \bm E\!\cdot\!\partial_{\bm k} f_{l'}^0
\right)~, \\
f_l^{\rm sk4,(1)} &= \frac{e\tau^2}{\hbar} \sum_{l'} w^{(4),\rm A}_{ll'}
\left( \bm E\!\cdot\!\partial_{\bm k} f_l^0 + \bm E\!\cdot\!\partial_{\bm k} f_{l'}^0 \right)~.
\end{align}
These results yield the linear-order nonequilibrium distribution functions that enter the calculation of the EOEE.
%
%
\section{Simplification of scattering rates}\label{Simplification_of_scattering_rates}
%
%
Consistent with the collision-integral convention, $w_{ll'}$ describes transitions from the state $\ket{l'}$ to $\ket{l}$. The rate is determined by Fermi's golden rule,
\be\label{ScRt}
w_{ll'} =\frac{2\pi}{\hbar}  
\bigg \langle 
|\bra{l} V_{\rm imp} \ket{l'_{\rm dis}}|^{2}
\bigg \rangle_{\text{dis}} 
\delta(\varepsilon_l - \varepsilon_{l'})~.
\ee
In the following, we simplify the scattering rate by considering different orders of the impurity potential. We use the approximations described below to obtain computationally tractable expressions for $w_{ll'}$. Expanding the Lippmann--Schwinger equation up to second order in the impurity potential, the eigenstate of the full Hamiltonian,
$H=\mathcal{H}+V_{\rm imp}$, can be written as
\bea
\ket{l'_{\rm dis}} &=& \ket{l'}+(\varepsilon_{l'} - \mathcal{H} +i\eta)^{-1} [ \hat{V}_{\rm imp}+\hat{V}_{\rm imp} (\varepsilon_{l'} - \mathcal{H} +i\eta)^{-1} \hat{V}_{\rm imp} ] \ket{l'}\nn\\
&=& \ket{l'}+\sum_{l''}\frac{V_{l''l'}}{(\varepsilon_{l'}-\varepsilon_{l''}+i\eta)}\ket{l''}+\sum_{l''l'''}\frac{V_{l''l'''}V_{l'''l'}}{(\varepsilon_{l'}-\varepsilon_{l''}+i\eta)(\varepsilon_{l'}-\varepsilon_{l'''}+i\eta)}\ket{l''}~.\label{disorder_perturbed_state}
\eea
Substituting the disorder-corrected eigenstate into Eq.~\eqref{ScRt} and expanding the scattering rate in powers of the impurity potential, we obtain
\begin{align}
    w_{ll'} = w_{ll'}^{(2)} + w_{ll'}^{(3)} + w_{ll'}^{(4)} + \cdots~,
\end{align}
where $w_{ll'}^{(n)} \propto V^n$ denotes the $n$th-order contribution to the scattering rate $(n=2,3,4,\ldots)$. The explicit expressions for these scattering rates are given by
\bea
\label{scrt2_zero_field}
w_{ll'}^{(2)} &=& \frac{2\pi}{\hbar} \left \langle V_{ll'}V_{l'l} \right \rangle_{\text{dis}} \delta(\varepsilon_l - \varepsilon_{l'})~,\\
\label{scat3}
w_{ll'}^{(3)} &=& \frac{2\pi}{\hbar} \sum_{l''}\bigg[\frac{\langle V_{ll''}V_{l''l'}V_{l'l} \rangle_{\text{dis}}}{\varepsilon_{l'} - \varepsilon_{l''}+i\eta} +\frac{\langle V_{ll'}V_{l'l''}V_{l''l} \rangle_{\text{dis}}}{\varepsilon_{l'} - \varepsilon_{l''} -i\eta}\bigg]\delta(\varepsilon_l - \varepsilon_{l'})~,\\
\label{scat4}
w_{ll'}^{(4)} &=& \frac{2\pi}{\hbar} \sum_{l''l'''}\bigg[\frac{\langle 
V_{ll'''}V_{l'''l'}V_{l'l''}V_{l''l} \rangle_{\text{dis}}}{(\varepsilon_{l'} - \varepsilon_{l''}-i\eta)(\varepsilon_{l'} - \varepsilon_{l'''}+i\eta)}+\frac{\langle V_{ll''}V_{l''l'''}V_{l'''l'}V_{l'l}
\rangle_{\text{dis}}}{(\varepsilon_{l'} - \varepsilon_{l''}+i\eta)(\varepsilon_{l'} - \varepsilon_{l'''}+i\eta)}\nn\\&&+\frac{\langle 
V_{ll'}V_{l'l'''}V_{l'''l''}V_{l''l} \rangle_{\text{dis}}}{(\varepsilon_{l'} - \varepsilon_{l''}-i\eta)(\varepsilon_{l'} - \varepsilon_{l'''}-i\eta)}\bigg]\delta(\varepsilon_l - \varepsilon_{l'})~. 
\eea
The second-order scattering rate, $w_{ll'}^{(2)}$, is symmetric under the exchange of the initial and final states, $l \leftrightarrow l'$. However, the higher-order scattering rates, $w_{ll'}^{(3)}$ and $w_{ll'}^{(4)}$, generally do not satisfy this symmetry. The scattering rates can therefore be decomposed as
\begin{align}
w_{ll'}^{(n),\rm S}=\frac{w_{ll'}^{(n)}+w_{l'l}^{(n)}}{2}~,~~~w_{ll'}^{(n),\rm A}=\frac{w_{ll'}^{(n)}-w_{l'l}^{(n)}}{2}~.
\end{align}
In the weak-disorder truncation used here, the symmetric parts of the third- and fourth-order rates give higher-order corrections to the conventional relaxation kernel. We absorb them into the symmetric rate and retain the antisymmetric parts, which generate the skew-scattering channel.
\subsection{Symmetric second-order scattering rate}
The second-order scattering rate, given in Eq.~\eqref{scrt2_zero_field}, was derived in the absence of an external electric field. When the electric field is taken into account, an electron acquires an additional energy due to the work done by the field as it undergoes a coordinate shift within the unit cell during a scattering event. Consequently, the second-order scattering rate is modified as
\begin{align}
\label{scrt2}
w_{l'l}^{(2)} = \frac{2\pi}{\hbar} \left \langle V_{l'l}V_{ll'} \right \rangle_{\text{dis}} \delta(\varepsilon_l - \varepsilon_{l'}-e{\bm E}\cdot \delta {\bm r}_{l'l})~,
\end{align}
where $\delta\bm{r}_{l'l}$ denotes the coordinate shift defined in Eq.~\eqref{appx:coordinate_shift}. In the present work, we retain only the electric-field-induced correction to the second-order scattering rate and neglect mixed contributions involving skew scattering and coordinate shifts, since they are subleading in the weak-disorder limit. Accordingly, the second-order scattering rate is decomposed into a field-independent symmetric contribution and an electric-field-induced coordinate-shift correction. The field-independent symmetric contribution governs the conventional momentum-relaxation process, the coordinate-shift correction gives rise to the side-jump contribution, while the antisymmetric scattering rates describe asymmetric scattering processes responsible for the skew-scattering contribution to the non-equilibrium distribution function.

For further simplification, we begin with the symmetric second-order scattering rate, $w_{ll'}^{(2),\mathrm{S}}$,
\bea
w_{ll'}^{(2), \rm S} &=& \frac{2\pi}{\hbar} \left \langle 
V_{ll'}V_{l'l} \right \rangle_{\text{dis}} \delta(\varepsilon_l - \varepsilon_{l'})=\frac{2\pi}{\hbar} \left \langle V_{\kb\kb'}V_{\kb'\kb}
\right \rangle_{\text{dis}} u^{\bm k \bm k'}_{n n'} u^{\bm k' \bm k}_{n' n} 
\delta(\varepsilon_{ n \bm k} - \varepsilon_{ n'\bm k'})~.
\eea
Using the disorder average,
$\langle V_{\bm{k}\bm{k}'}V_{\bm{k}'\bm{k}}\rangle_{\rm dis}
= n_iV_0^2$,
together with the assumption that only intraband scattering is allowed ($n'=n$), the expression for the symmetric second-order scattering rate, $w_{ll'}^{(2),\mathrm{S}}$, reduces to
\be\label{ScRt2}
w_{n,\kb\kb'}^{(2), \rm S}=\frac{2\pi n_i V_0^2}{\hbar} | u^{\bm k \bm k'}_n |^2 \delta(\varepsilon_{ n \bm k} - \varepsilon_{ n \bm k'})~.
\ee
To further simplify the scattering rate, we adopt the small-momentum-transfer approximation,
$\bm q=\bm{k}'-\bm{k}\rightarrow0$.
In this limit, the cell-periodic part of the Bloch state can be expanded about $\bm{k}$ as
\be\label{expand_u}
\ket{u_{ n \kb'}}=\ket{u_{ n \bm{k}}}+q_b \ket{\partial_b u_{ n \bm{k}}}+\frac{1}{2}q_b q_c\ket{\partial_b \partial_c u_{ n \bm{k}}}+\cdots~.
\ee
With this expansion, the overlap matrix of the cell-periodic parts of the Bloch states becomes
\be\label{expand_uu}
u^{\bm k \bm k'}_n \approx 1-iq_b\mathcal{R}_{ n n}^{b}(\bm k)\approx e^{-iq_b\mathcal{R}^b_{ n n}(\bm k)}~. 
\ee
Substituting this into Eq.~\eqref{ScRt2} gives
\be\label{final_SR2}
   w_{n,\kb\kb'}^{(2), \rm S}=\frac{2\pi n_i V_0^2}{\hbar} |u^{\bm k \bm k'}_n|^2 \delta(\varepsilon_{ n \bm k} - \varepsilon_{ n \bm k'})\approx \frac{2\pi n_i V_0^2}{\hbar} \delta(\varepsilon_{ n \bm k} - \varepsilon_{ n \bm k'})~.
\ee
\subsection{Antisymmetric third-order scattering rate}\label{app_third_order_scattering_rates}
The antisymmetric part of the third-order scattering rate is given by
\bea
w_{ll'}^{(3),\rm A} &=& \frac{1}{2}(w_{ll'}^{(3)}-w_{l'l}^{(3)}) \nn\\
&=& \frac{\pi}{\hbar} \sum_{l''}\bigg[\frac{\langle V_{ll''}V_{l''l'}V_{l'l}
\rangle_{\text{dis}}}{\varepsilon_{l} - \varepsilon_{l''}+i\eta} 
+ \frac{\langle V_{ll'}V_{l'l''}V_{l''l} \rangle_{\text{dis}}}{\varepsilon_{l} - \varepsilon_{l''}-i\eta} -\frac{\langle V_{l'l''}V_{l''l}V_{ll'}
\rangle_{\text{dis}}}{\varepsilon_{l} - \varepsilon_{l''}+i\eta} 
-\frac{\langle V_{l'l}V_{ll''}V_{l''l'} \rangle_{\text{dis}}}{\varepsilon_{l} - \varepsilon_{l''}-i\eta}\bigg]\delta(\varepsilon_l - \varepsilon_{l'})\nn\\
&=& \frac{\pi}{\hbar} \sum_{l''}\bigg[\langle 
V_{ll''}V_{l''l'}V_{l'l}\rangle_{\rm dis} \left (\frac{1}{\varepsilon_{l} - \varepsilon_{l''}+i\eta}-\frac{1}{\varepsilon_{l} - \varepsilon_{l''}-i\eta} \right) - \text{c.c.} \bigg]\delta(\varepsilon_l - \varepsilon_{l'})\nn\\
&=& \frac{4 \pi^2}{\hbar} {\rm Im}\sum_{l''}\langle V_{ll''}V_{l''l'}V_{l'l}
\rangle_{\rm dis}\delta(\varepsilon_{l} - \varepsilon_{l'})\delta(\varepsilon_{l} - \varepsilon_{l''})\nn\\
&=& \frac{4 \pi^2}{\hbar} {\rm Im}\sum_{n''} \sum_{ \bm k''}\langle V_{\kb\kb''}^0 V_{\kb''\kb'}^0 V_{\kb'\kb}^0
\rangle_{\rm dis}~u^{\bm k \bm k''}_{n n''} u^{\bm k'' \bm k'}_{n'' n'} u^{\bm k' \bm k}_{n' n} \delta(\varepsilon_{ n \bm k} - \varepsilon_{ n' \bm k'})\delta(\varepsilon_{ n\bm k} - \varepsilon_{ n'' \bm k''})~.
\eea
To obtain a more tractable form of this expression, we impose the same approximations adopted in the previous analysis. We consider the case where the Fermi surface is formed by a single band and restrict the impurity-induced transitions to intraband scattering processes, such that the initial and final states have an identical band index $n$ while differing in their crystal momenta $\bm{k}$~\cite{Cong_Xiao_24_prb, Harsh_2026_prb, sanjay_2026_prb}. As a consequence, all intermediate band summations collapse to a single band, yielding the constraint $n''=n'=n$. The disorder-averaged third-order impurity correlator takes the form
\begin{align}
\langle V_{\kb\kb''}^0 V_{\kb''\kb'}^0 V_{\kb'\kb}^0\rangle_{\rm dis}
&=\sum_{i,j,k}V_i V_j V_k \exp\left[i(\kb''-\bm{k})\cdot\bm R_i+i(\kb'-\kb'')\cdot\bm R_j+i(\bm{k}-\kb')\cdot\bm R_k\right]=n_iV_1^3~.
\end{align}
where $n_i$ is the impurity concentration and $V_1$ is the first moment of the impurity potential, with units of energy times area for the two-dimensional disorder model. The antisymmetric third-order scattering rate then takes the compact form
\bea\label{third_order_sct}
w_{n,\kb\kb'}^{(3),\rm A} = \frac{4\pi^2 n_i V_1^3}{\hbar}{\rm Im}\sum_{\bm k''} u^{\bm k \bm k''}_{n} u^{\bm k'' \bm k'}_{n} u^{\bm k' \bm k}_{n} \delta(\varepsilon_{ n \bm k} - \varepsilon_{ n \bm k'})\delta(\varepsilon_{ n\bm k} - \varepsilon_{ n \bm k''})~.
\eea
For further simplification of the antisymmetric third-order scattering rate, we evaluate the overlap matrix elements of the cell-periodic parts of the Bloch states. In addition to $\bm q \to 0$, we assume $q'=(\kb''-\bm{k})\to0$, which gives $\ket{u_{ n \kb''}}=\ket{u_{ n \bm{k}}}+q_b' \ket{\partial_b u_{ n \bm{k}}}+\frac{1}{2}q_b' q_c'\ket{\partial_b \partial_c u_{ n \bm{k}}}+\cdots$. As a result, the overlaps are
\bea
    u^{\bm k \bm k''}_n &=&1-iq_b'\mathcal{R}^b_{ n n}+\frac{1}{2}q_b'q_c'\langle u_{ n \bm k}|\partial_b \partial_c u_{ n \bm k}\rangle~,\nn\\
    u^{\bm k'' \bm k'}_n &=& 1+i(q_b'-q_b)\mathcal{R}_{ n n}^b +q_b q_c'\langle \partial_c u_{ n \bm{k}}|\partial_b u_{ n \bm{k}}\rangle+\frac{1}{2}q_b q_c \langle u_{ n \bm k}|\partial_b \partial_c u_{ n \bm k}\rangle+\frac{1}{2}q_b' q_c'\langle \partial_b \partial_c u_{ n \bm k}|u_{ n \bm k}\rangle~,\nn\\
    u^{\bm k' \bm k}_n &=& 1+iq_b\mathcal{R}^b_{ n n}+\frac{1}{2}q_b q_c\langle \partial_b \partial_c u_{ n \bm k}| u_{ n \bm k}\rangle~. \nn
\eea
Using these overlaps in Eq.~\eqref{third_order_sct}, we can simplify $w_{n,\kb\kb'}^{(3),A}$ as follows:
\bea\label{w_3a_final}
    w_{n,\kb\kb'}^{(3),\rm A} &=& \frac{4\pi^2 n_i V_1^3}{\hbar}{\rm Im}\sum_{\bm k''} u^{\bm k \bm k''}_n u^{\bm k'' \bm k'}_n u^{\bm k' \bm k}_n \delta(\varepsilon_{ n \bm k} - \varepsilon_{ n \bm k'})\delta(\varepsilon_{ n\bm k} - \varepsilon_{ n \bm k''}) \nn\\
    &=& \frac{4\pi^2 n_i V_1^3}{\hbar}\sum_{\bm k''}{\rm Im}[1+(q_bq_c+q_b'q_c'-q_bq_c')\mathcal{R}^b_{ n n}\mathcal{R}^c_{ n n}+q_bq_c'\langle \partial_c u_{ n\bm{k}}|\partial_b u_{ n\bm{k}}\rangle \nn \\&&\qquad\qquad\qquad +(q_b q_c+q_b' q_c'){\rm Re}(\langle u_{ n \bm{k}}|\partial_b\partial_c u_{ n \bm{k}}\rangle)]\delta(\varepsilon_{ n \bm k}- \varepsilon_{ n \bm k'})\delta(\varepsilon_{ n\bm k} - \varepsilon_{ n \bm k''}) \nn\\
    &=& \frac{4\pi^2 n_i V_1^3}{\hbar}\sum_{ \kb''}q_b q_c'{\rm Im}[\langle \partial_c u_{ n\bm{k}}|\partial_b u_{ n\bm{k}}\rangle]\delta(\varepsilon_{ n \bm k}- \varepsilon_{ n \bm k'})\delta(\varepsilon_{ n\bm k} - \varepsilon_{ n \bm k''})\nn\\
    &=& \frac{2\pi^2 n_i V_1^3}{\hbar}\sum_{ \kb''}(k_b'-k_b) (k_c''-k_c)\Omega^{bc}_ n(\bm{k}) \delta(\varepsilon_{ n \bm k}- \varepsilon_{ n \bm k'})\delta(\varepsilon_{ n\bm k} - \varepsilon_{ n \bm k''}) \nn\\
    &=& \frac{2\pi^2 n_i V_1^3}{\hbar}\sum_{ \kb''}[(k_b'-k_b) (k_c''-k_c)]\epsilon_{bcd}\Omega^{d}_ n(\bm{k}) \delta(\varepsilon_{ n \bm k}- \varepsilon_{ n \bm k'})\delta(\varepsilon_{ n\bm k} - \varepsilon_{ n \bm k''}) \nn\\
    &=& \frac{2\pi^2 n_i V_1^3}{\hbar}\sum_{ \kb''}[(\kb'-\bm{k})\times(\kb''-\bm{k})]\cdot\bm{\Omega}_ n(\bm{k}) \delta(\varepsilon_{ n \bm k}- \varepsilon_{ n \bm k'})\delta(\varepsilon_{ n\bm k} - \varepsilon_{ n \kb''}) \nn\\
    &=& -\frac{2\pi^2 n_i V_1^3}{\hbar}\sum_{ \kb''}[(\kb''\times\kb')+(\kb'\times\bm{k})+(\bm{k}\times\kb'')]\cdot\bm{\Omega}_ n(\bm{k}) \delta(\varepsilon_{ n \bm k}- \varepsilon_{ n \bm k'})\delta(\varepsilon_{ n\bm k} - \varepsilon_{ n \bm k''})~.   
\eea
Here, $\Omega_{ n}^{bc}$ is the BC of band $n$ and is defined as $\Omega_{ n}^{bc}=-\Omega_{ n}^{cb}=2\mathrm{Im}[\langle\partial_c u_{n\bm k}|\partial_b u_{n\bm k}\rangle]$.
\subsection{Antisymmetric fourth-order scattering rate}\label{app_fourth_order_scattering_rates}
Following the procedure used for $w_{ll'}^{(3), \rm A}$, we calculate $w^{(4),A}_{ll'}$ as
\bea
   w^{(4),\rm A}_{ll'} &=& -\dfrac{4\pi^2}{\hbar}\sum_{l''l'''}{\rm Im}\left [\langle V_{ll'''}V_{l'''l'}V_{l'l''}V_{l''l}\rangle_{\rm dis}-\langle V_{ll''}V_{l''l'''}V_{l'''l'}V_{l'l}\rangle_{\rm dis}\right.\nn\\
   &&\left.{}-\langle V_{ll'''}V_{l'''l''}V_{l''l'}V_{l'l}\rangle_{\rm dis}\right ]
   \dfrac{\delta(\varepsilon_l-\varepsilon_{l'})\,\delta(\varepsilon_{l'}-\varepsilon_{l''})}{(\varepsilon_{l'}-\varepsilon_{l'''})}~.
\eea
The structure of this scattering rate is qualitatively different from the cases considered previously. In particular, restricting all scattering processes to be purely intraband, as assumed earlier, leads to a vanishing contribution, as shown below. We therefore relax the intraband-scattering constraint for the intermediate state and consider
$l,l',l''=(n,\bm{k}),(n,\bm{k'}),(n,\bm{k''})$, while
$l'''=(n',\bm{k'''})$, where the intermediate band index $n'$ is allowed to differ from the initial band index $n$. With this modified approximation, the expression simplifies to
\bea\label{sct4}
   w^{(4),\rm A}_{n,\kb\kb'} &=& -\dfrac{4\pi^2}{\hbar}\sum_{ n'}\sum_{\kb''\kb'''}{\rm Im}\bigg[\langle V_{\kb\kb'''}V_{\kb'''\kb'}V_{\kb'\kb''}V_{\kb''\kb}\rangle_{\rm dis} u^{\bm k \bm k'''}_{n n'} u^{\bm k''' \bm k'}_{n' n} u^{\bm k' \bm k''}_n u^{\bm k'' \bm k}_n  -\langle V_{\kb\kb''}V_{\kb''\kb'''}V_{\kb'''\kb'}V_{\kb'\kb}\rangle_{\rm dis} \nn \\ &\quad& \times u^{\bm k \bm k''}_n u^{\bm k'' \bm k'''}_{n n'} u^{\bm k''' \bm k'}_{n' n} u^{\bm k' \bm k}_n  -\langle V_{\kb\kb'''}V_{\kb'''\kb''}V_{\kb''\kb'}V_{\kb'\kb}\rangle_{\rm dis} u^{\bm k \bm k'''}_{n n'} u^{\bm k''' \bm k''}_{n' n} u^{\bm k'' \bm k'}_n u^{\bm k' \bm k}_n \bigg]\dfrac{\delta(\varepsilon_{ n \bm{k}}-\varepsilon_{n\bm k'})\delta(\varepsilon_{ n\kb'}-\varepsilon_{ n\kb''})}{(\varepsilon_{ n\kb'}-\varepsilon_{ n'\kb'''})}\nn\\
   &=&-\dfrac{4\pi^2 n_i^2 V_0^4}{\hbar}\sum_{ n'}\sum_{\kb''\kb'''}{\rm Im}\bigg[u^{\bm k \bm k'''}_{n n'} u^{\bm k''' \bm k'}_{n' n} u^{\bm k' \bm k''}_n u^{\bm k'' \bm k}_n -u^{\bm k \bm k''}_n u^{\bm k'' \bm k'''}_{n n'} u^{\bm k''' \bm k'}_{n' n} u^{\bm k' \bm k}_n -u^{\bm k \bm k'''}_{n n'} u^{\bm k''' \bm k''}_{n' n} u^{\bm k'' \bm k'}_n u^{\bm k' \bm k}_n \bigg] \nn \\ &\quad& \times  \dfrac{\delta(\varepsilon_{ n \bm{k}}-\varepsilon_{n\bm k'})\delta(\varepsilon_{ n\kb'}-\varepsilon_{ n\kb''})}{(\varepsilon_{ n\kb'}-\varepsilon_{ n'\kb'''})}~. \nn \\ 
\eea
Here, we have used Wick's theorem to reduce the disorder average of the four-point impurity correlator into a product of two-point correlators. For Gaussian-distributed disorder, this factorization takes the form
$
\left\langle V_{\bm{k}\bm{k}''} V_{\bm{k}''\bm{k}'} V_{\bm{k}'\bm{k}'''} V_{\bm{k}'''\bm{k}} \right\rangle_{\rm dis} = \left\langle V_{\bm{k}\bm{k}''} V_{\bm{k}''\bm{k}'} \right\rangle_{\rm dis}
\left\langle V_{\bm{k}'\bm{k}'''} V_{\bm{k}'''\bm{k}} \right\rangle_{\rm dis} = n_i^2 V_0^4.
$
Using the same small-momentum-transfer approximation, together with $q''=(\kb'''-\bm{k})\to 0$, we get 
\bea
{\rm Im}[u^{\bm k \bm k'''}_{n n'} u^{\bm k''' \bm k'}_{n' n} u^{\bm k' \bm k''}_n u^{\bm k'' \bm k}_n] &=& {\rm Im}[q_b''(q_c''-q_c)\mathcal{R}^{b}_{nn'}\mathcal{R}^{c}_{n'n}]=-\frac{1}{2}q_b''(q_c''-q_c)\Omega^{bc}_{ n n'}~,\nn\\
{\rm Im}[u^{\bm k \bm k''}_{n} u^{\bm k'' \bm k'''}_{n n'} u^{\bm k''' \bm k'}_{n' n} u^{\bm k' \bm k}_n] &=& {\rm Im}[(q_b''-q_b')(q_c''-q_c)\mathcal{R}^{b}_{nn'}\mathcal{R}^{c}_{n'n}]=-\frac{1}{2}(q_b''-q_b')(q_c''-q_c)\Omega^{bc}_{ n n'}~,\nn\\
{\rm Im}[u^{\bm k \bm k'''}_{n n'} u^{\bm k''' \bm k''}_{n' n} u^{\bm k'' \bm k'}_n u^{\bm k' \bm k}_n] &=& {\rm Im}[q_b''(q_c''-q_c')\mathcal{R}^{b}_{nn'}\mathcal{R}^c_{n'n}]=-\frac{1}{2}q_b''(q_c''-q_c')\Omega^{bc}_{ n n'}~.\nn
\eea
For $n'=n$, the product
$\mathcal{R}^{b}_{nn'}\mathcal{R}^{c}_{n'n}$ becomes purely real. Hence, the imaginary part of this product vanishes, eliminating the corresponding intraband contribution to the scattering rate. This demonstrates that only interband virtual transitions, characterized by $n'\neq n$, contribute to the finite scattering rate. Substituting this condition into Eq.~\eqref{sct4}, we obtain
\bea
w^{(4),\rm A}_{n,\kb\kb'} &=& \dfrac{2\pi^2 n_i^2 V_0^4}{\hbar}\sum_{n'\neq n}\sum_{\kb''\kb'''}\bigg[q_b''(q_c''-q_c)-(q_b''-q_b')(q_c''-q_c)-q_b''(q_c''-q_c')\bigg]\Omega^{bc}_{ n n'}\dfrac{\delta(\varepsilon_{ n \bm{k}}-\varepsilon_{ n \kb'})\delta(\varepsilon_{ n\kb'}-\varepsilon_{ n\kb''})}{(\varepsilon_{ n\kb'}-\varepsilon_{ n'\kb'''})} \nn\\
&=& \dfrac{2\pi^2 n_i^2 V_0^4}{\hbar}\sum_{n'\neq n}\sum_{\kb''\kb'''} \epsilon_{bcd} \bigg[(k_b''-k_b)(k_c'''-k_c')-(k_c'''-k_c'')(k_b'''-k_b)\bigg]\Omega^{d}_{ n n'}\dfrac{\delta(\varepsilon_{ n \bm{k}}-\varepsilon_{n \kb'})\delta(\varepsilon_{ n\kb'}-\varepsilon_{ n\kb''})}{(\varepsilon_{ n\kb'}-\varepsilon_{ n'\kb'''})} \nn\\
&=& \dfrac{2\pi^2 n_i^2 V_0^4}{\hbar}\sum_{n'\neq n}\sum_{\kb''\kb'''}\bigg[(\kb''-\bm{k})\times(\kb'''-\kb')-(\kb'''-\bm{k})\times(\kb'''-\kb'')\bigg]\cdot\bm{\Omega}_{ n n'}\dfrac{\delta(\varepsilon_{ n \bm{k}}-\varepsilon_{ n \kb'})\delta(\varepsilon_{ n\kb'}-\varepsilon_{ n\kb''})}{(\varepsilon_{ n\kb'}-\varepsilon_{ n'\kb'''})} \nn\\
&=& -\dfrac{2\pi^2 n_i^2 V_0^4}{\hbar}\sum_{n'\neq n}\sum_{\kb''\kb'''}\bigg[(\kb'\times\bm{k})+(\kb''\times\kb')+(\bm{k}\times\kb'')\bigg]\cdot\bm{\Omega}_{ n n'}\dfrac{\delta(\varepsilon_{ n \bm{k}}-\varepsilon_{n \kb'})\delta(\varepsilon_{ n\kb'}-\varepsilon_{ n\kb''})}{(\varepsilon_{ n\kb'}-\varepsilon_{ n'\kb'''})}~.
\eea
Here, $\Omega^{bc}_{nn'}=-2\mathrm{Im}\left[ \mathcal{R}^{b}_{nn'}\mathcal{R}^{c}_{n'n}\right]$ denotes the interband geometric quantity associated with the virtual transition between bands $n$ and $n'$. In addition, we simplify the remaining momentum summation by taking $\sum_{\bm{k}'''}
(\varepsilon_{n\bm{k}'}-\varepsilon_{n'\bm{k}'''})^{-1} \approx (\varepsilon_{n\bm{k}'}-\varepsilon_{n'\bm{k}})^{-1}$, which corresponds to the approximation $\bm{k}'''\rightarrow\bm{k}$ within the Brillouin zone. Consequently, the scattering rate reduces to
\be\label{w_4a_final} w^{(4),A}_{n,\kb\kb'}=-\dfrac{2\pi^2 n_i^2 V_0^4 }{\hbar} \sum_{\kb''}\bigg[(\kb'\times\bm{k})+(\kb''\times\kb')+(\bm{k}\times\kb'')\bigg] \cdot \tilde{\bm{\Omega}}_n \delta(\varepsilon_{ n \bm{k}}-\varepsilon_{ n \bm {k'}})\delta(\varepsilon_{ n\bm{k}}-\varepsilon_{ n\kb''})~.
\ee
Here, $ \tilde{\bm\Omega}_n = \sum_{n'\neq n} \bm\Omega_{nn'}/(\varepsilon_{n\bm k} - \varepsilon_{n'\bm k}) $ is the energy-normalized BC for band $n$. Equation~\eqref{w_4a_final} is the Gaussian noncrossing fourth-order contribution obtained with $\bm k'''\to\bm k$ and used numerically.
\section{Simplification of the side-jump velocity}\label{side_jump_velocity}
The side-jump velocity defined in Eq. \eqref{sj_vel} is given by
\begin{align}
\bm v^{\rm sj}_l &= \sum_{l'} w_{l'l}^{(2),\rm S} \delta \bm r_{l'l} =- \sum_{l'} w_{l'l}^{(2),\rm S} \delta \bm r_{ll'} \nn\\
\bm v^{\rm sj}_{n\bm k} &=-  \frac{2\pi n_i V_0^2}{\hbar} \sum_{n'}\sum_{\kb'} |u_n^{\bm k\bm k'}|^2 \delta(\varepsilon_{ n \bm k} - \varepsilon_{ n \bm k'}) \delta {\bm r}_{nn}(\kb,\kb')~.
\end{align} 
To obtain the last line of the above equation, we use the expression for $w_{l'l}^{(2),\rm S} \simeq w_{n,\bm k\bm k'}^{(2),\rm S}$ from Eq.~\eqref{final_SR2}. The coordinate shift $\delta r^a_{nn}(\kb,\kb')$ during an elastic scattering event from state $\ket{n\bm k'}$ to state $\ket{n\bm k}$ is given by
\be\label{eq:pos_shift_app}
\delta r^a_{nn}(\kb, \kb') = R^a_{nn}(\kb)  - R^a_{nn}(\kb') - (\partial_a + \partial'_a) {\rm arg}(u_n^{\bm k\bm k'})~,
\ee
where the intraband Berry connection is $R^a_{nn}(\kb)
= \langle u_{n\kb} | i\partial_a u_{n\kb} \rangle$. Using the identity in Eq.~\eqref{expand_u}, we can write \( \partial'_a \vert u_{n\kb'} \rangle =  \ket{\partial_a u_{n \kb}} + q_b  \ket{ \partial_a \partial_b u_{n \kb}}\), which gives the Berry connection at $\bm k'$ in terms of $\mathcal{\bm R}_{nn} (\bm k)$ and the momentum-shift vector $\bm q$:
\be 
R^a_{nn}(\kb') = R^a_{nn}(\kb)  + i q_b \bigg[ \bra{u_{n\kb}} \partial_a\partial_b u_{n\kb} \rangle  + \bra{\partial_b u_{n\kb}} \partial_a u_{n\kb} \rangle\bigg]~. \nn
\ee 
Furthermore, the identity \( \bra{u_{n\kb}} \partial_a\partial_b u_{n\kb} \rangle = - \bra{\partial_a u_{n\kb}} \partial_b u_{n\kb} \rangle - i \partial_a R^b_{nn} \) gives
\be 
R^a_{nn}(\kb') = R^a_{nn}(\kb) - q_b \Omega^{ab}_n(\kb) +  q_b \partial_a R^b_{nn}(\kb)~, \nn
\ee 
where $\Omega^{ab}_n$ is the BC, given by
\begin{equation}
\Omega^{ab}_n(\kb) = i \left[ \langle \partial_a u_{n\kb} | \partial_b u_{n\kb} \rangle
- \langle \partial_b u_{n\kb} | \partial_a u_{n\kb} \rangle \right]~.
\end{equation}
To calculate the last term of the coordinate shift vector, we use Eq.~\eqref{expand_uu} for the overlap matrix between two Bloch states, which gives
\begin{align}
(\partial_{a} + \partial'_{a})\mathrm{arg} (u_n^{\bm k\bm k'}) &= (\partial_{a} + \partial'_{a}) \{- q_b R^b_{nn}(\kb)\} \nn\\ 
&= - q_b \partial_{a}\mathcal{R}^b_{nn} (\bm k)~.
\end{align}
Substituting all these expressions into Eq.~\eqref{eq:pos_shift_app}, the coordinate shift becomes
\begin{equation}
\delta r^a_{nn}(\kb,\kb')
= q_b \Omega^{ab}_n(\kb)~.
\end{equation}
Using the relation $\Omega^{ab}_n = \epsilon_{abc} \Omega^c_n$, the coordinate shift can be written more compactly as
\begin{equation}
\delta {\bm r}_{nn}(\kb,\kb')
= (\kb' - \kb) \times {\bm \Omega}_n(\kb)~.
\end{equation}
Combining this expression for the coordinate shift with the small-momentum-transfer approximation, $|u_n^{\bm k\bm k'}|^2 \approx 1$, gives the side-jump velocity,
\begin{align}
    \bm v^{\text{sj}}_{n\bm k} = \frac{2 \pi n_i V_0^2}{\hbar} \sum_{\bm k'} [ (\bm k - \bm k') \times \mathbf{\Omega}_n(\bm k) ] \delta(\varepsilon_{n\bm k} - \varepsilon_{n\bm k'})~.
\end{align}
%
%

    
\end{appendix}

\twocolumngrid

\bibliography{Ref}
\end{document}